\documentclass[12pt]{article}
\usepackage[dvips]{graphicx}
\usepackage{epsfig}
\psfigdriver{dvips}
\renewcommand{\thefootnote}{\fnsymbol{footnote}}

\newcommand{\prepr}[1] {\begin{flushright}  {\bf #1} \end{flushright} \vskip 1.cm}
\newcommand{\titul}[1] {\boldmath \begin{center}{\Large {\bf #1 } } \end{center}
\vskip 0.8cm}
\newcommand{\autor}[1] {\begin{center}  {\bf \lineskip .3cm #1  }
                        \end{center} }
\newcommand{\instit}[1] {\begin{center}  {\normalsize \bf \it #1   } \end{center}}
\newcounter{muni}

\def\bmaT{\left(\begin{array}{ccc}}
\def\emaT{\end{array}\right)}
\def\bma{\left( \begin{array} }
\def\ema{\end{array} \right)}
\def\gsim{~{\rlap{\lower 3.5pt\hbox{$\mathchar\sim$}}\raise 1pt\hbox{$>$}}\,}
\def\lsim{~{\rlap{\lower 3.5pt\hbox{$\mathchar\sim$}}\raise 1pt\hbox{$<$}}\,}
\def\beqa{\begin{eqnarray}}
\def\eeqa{\end{eqnarray}}
\def\beq{\begin{equation}}
\def\eeq{\end{equation}}
\def\nn{\nonumber}
\begin{document}
\hbadness=10000
\pagenumbering{arabic}
\begin{titlepage}
\prepr{hep-ph/yymmnnn\\
\hspace{30mm} KEK--TH--1106\\
\hspace{30mm} October 2006}

\begin{center}
\titul{\bf Lepton Flavour Violating $\tau$ Decays in the\\
Left-Right Symmetric Model}

\autor{A.G. Akeroyd$^{1,3,4}$\footnote{akeroyd@mail.ncku.edu.tw}, 
Mayumi Aoki$^1$\footnote{mayumi.aoki@kek.jp} 
and Yasuhiro Okada$^{1,2}$\footnote{yasuhiro.okada@kek.jp}}
\instit{1: Theory Group, KEK, 1-1 Oho, \\
Tsukuba, Ibaraki, 305-0801 Japan}
\instit{2: The Graduate University for Advanced Studies (Sokendai),\\
1-1 Oho, Tsukuba, Ibaraki, 305-0801 Japan
}
\instit{3: Department of Physics,\\ 
National Cheng Kung University,
Tainan, 701 Taiwan
}
\instit{4: National Center for Theoretical Sciences,\\
Taiwan \\
}

\end{center}

\vskip0.5cm

\begin{abstract}
\noindent
The Left-Right symmetric extension of the
Standard Model with Higgs isospin triplets 
can provide neutrino masses via a TeV scale seesaw mechanism.
The doubly charged Higgs bosons 
$H^{\pm\pm}_L$ and $H^{\pm\pm}_R$ induce
lepton flavour violating decays $\tau^\pm \to lll$ at tree-level
via a coupling which is related to the 
Maki-Nakagawa-Sakata matrix $(V_{\rm MNS})$. 
We study the magnitude and correlation of $\tau^\pm \to lll$ 
and $\mu\to e\gamma$ with specific assumptions for the
origin of the large mixing in $V_{\rm MNS}$ while respecting the 
stringent bound for $\mu\to eee$.
It is also shown that an angular asymmetry for
$\tau^\pm \to lll$ is sensitive to the relative strength of 
the $H^{\pm\pm}_L$ and $H^{\pm\pm}_R$ mediated contributions
and provides a means of distinguishing models with 
doubly charged Higgs bosons.
\end{abstract}
\vskip1.0cm
{\bf  PACS index: 13.35.-r, 14.60.Pq, 14.80.Cp} 
\end{titlepage}
\thispagestyle{empty}
\newpage

\pagestyle{plain}
\renewcommand{\thefootnote}{\arabic{footnote} }
\setcounter{footnote}{0}

\section{Introduction}
In recent years there has been increasing evidence 
that neutrinos oscillate and possess a small mass below the eV scale
\cite{Fukuda:1998mi}.
This revelation necessitates physics beyond the Standard Model
(SM), which could 
manifest itself at the CERN Large Hadron Collider (LHC) 
and/or in low energy experiments which 
search for lepton flavour violation (LFV) \cite{Kuno:1999jp}.
Consequently, models of neutrino mass generation which can be probed
at present and forthcoming experiments are of great phenomenological
interest. 

Massive neutrinos may be accommodated by adding a 
$SU(2)_L$ singlet (``sterile'') right-handed neutrino $\nu_R$ to the SM
Lagrangian together with the corresponding Dirac mass term.
In order to obtain masses of the eV scale,
the Yukawa coupling of the neutrinos to the SM Higgs
boson would need to be at least 6 orders
of magnitude smaller than the electron Yukawa coupling.
Moreover, there would be no observable phenomenological consequences
aside from neutrino oscillations.
More appealing frameworks for neutrino mass generation can 
be found if neutrinos are of the Majorana type. 
The celebrated seesaw mechanism \cite{Minkowski:1977sc}
ascribes the smallness of the
neutrino mass to the large scale of the unobserved
{\it heavy} right-handed neutrinos ($N_R$).  
Dirac mass terms of the order
of the top quark mass would require $N_R\sim 10^{14}$ GeV,
a scale which is far beyond the reach of any envisioned 
collider.
Reducing the scale of $N_R$ to the order of a few TeV would require
Dirac mass terms of the order of MeV, which constitutes a mild
fine-tuning with respect to magnitude of the charged lepton masses.
However, such a choice would permit the
mechanism to be probed at future high-energy colliders.
This ``low energy seesaw mechanism'' may be implemented in
Left-Right (LR) symmetric models \cite{Pati:1974yy} 
with Higgs triplet representations \cite{Mohapatra:1979ia},
in which the mass matrix for $N_R$ is given
by the product of a new Yukawa coupling $h_{ij}$ and 
a triplet vacuum expectation value (vev) $v_R$. 
The mass scale of the other new particles
in the model (e.g. the new gauge bosons $W_R,Z_R$ and doubly charged 
Higgs bosons $H^{\pm\pm}_{L,R}$) is also determined by
$v_R$, resulting in a rich phenomenology
at the LHC if  $v_R\sim {\rm TeV}$ \cite{Gunion:1989in},\cite{Deshpande:1990ip}.

The Yukawa coupling $h_{ij}$ mediates many low energy LFV processes. 
In this paper we consider the impact of 
$H^{\pm\pm}_{L,R}$ on the branching ratio (BR) of the LFV decays
$\tau\to l_i l_j l_k$ and $\mu\to e\gamma$ in the context of the 
LR symmetric model \cite{Lim:1981kv},\cite{Cirigliano:2004mv}.
Experimental prospects for $\mu\to e\gamma$ are bright with the
imminent commencement of the MEG experiment which will
probe BR$\sim 10^{-13}\to 10^{-14}$, two to 
three orders of magnitude beyond the current upper limit
\cite{Grassi:2005ac}.
At the $e^+e^-$ B factories  limits of the order
BR($\tau\to l_i l_j l_k) <10^{-7}$ with $\sim 90$ fb$^{-1}$
\cite{Yusa:2004gm},\cite{Aubert:2003pc} have been obtained
utilizing direct $e^+e^-\to \tau^+\tau^-$ production.
Simulations of the detection prospects
at a proposed high luminosity $e^+e^-$ B factory with 
${\cal L}$=5$\to 50$ ab$^{-1}$ anticipate sensitivity to
BR$\sim 10^{-8}\to 10^{-9}$ \cite{Hashimoto:2004sm}.
Additional searches can be performed at
the LHC where $\tau$ leptons are copiously produced
from the decays of $W,Z,B,D$, with anticipated sensitivities to 
BR$\sim 10^{-8}$ \cite{Santinelli:2002ea}.

In the LR symmetric model $H^{\pm\pm}_{L,R}$ mediate $\tau\to l_i l_j l_k$
at {\it tree-level} due to an effective 4-Fermi charged lepton 
interaction proportional to $h_{\tau i}^\ast h_{jk}/M^2_{H^{\pm\pm}}$.
Hence such a model can comfortably accommodate BRs of
order $10^{-7}\to 10^{-9}$ which will be probed
at current and forthcoming experiments. 
For the loop induced decays $\mu\to e\gamma$ and
$\tau\to l\gamma$ the dominant
contribution in the LR symmetric model 
originates from diagrams involving $H^{\pm\pm}_{L,R}$.
In the LR model one has BR($\tau\to lll)\gg$
BR$(\tau\to l\gamma)$, which contrasts
with the general expectation 
BR$(\tau\to l\gamma)\gg $BR($\tau\to lll$) for models
in which the tree-level $\tau\to l_il_jl_k$ interaction
is absent (for scenarios where BR$(\tau\to l\gamma)\sim $BR($\tau\to lll$) is
possible see \cite{Dedes:2002rh}). 
Due to the larger backgrounds for the search for
$\tau\to l\gamma$, the experimental sensitivity to
BR$(\tau\to l\gamma)$ at the $e^+e^-$ B factories is expected
to be inferior to that for $\tau\to l_il_jl_k$ 
\cite{Hashimoto:2004sm}.
Consequently the hierarchy BR($\tau\to lll)\gg$ 
BR$(\tau\to l\gamma)$ in the LR model affords 
more promising detection prospects.

The presence of the above tree-level 4-Fermi 
interaction would also mediate the decay $\mu\to eee$ for which there is
a strict bound ($< 10^{-12}$ \cite{Bellgardt:1987du}) 
at least three orders of magnitude
stronger than the anticipated experimental sensitivity to $\tau\to lll$. 
Hence obtaining BR($\tau\to lll)> 10^{-9}$ together
with compliance of the above bound on BR($\mu\to eee$) 
restricts the structure of $h_{ij}$.
In the LR model we consider a specific ansatz for $h_{ij}$
motivated by the observed pattern of the neutrino mixing angles,
and perform a numerical study of the magnitude of
BR$(\tau\to l_il_jl_k)$.

Observation of BR($\tau\to lll)> 10^{-9}$ would be a spectacular
signal of LFV and could be readily accommodated by a 
tree-level 4-Fermi coupling
such as $h_{\tau i}^\ast h_{jk}/M^2_{H^{\pm\pm}}$ in the LR model.
However, other models which contain a $H^{\pm\pm}_L$ or 
$H^{\pm\pm}_R$ with an analogous $h_{ij}$ leptonic
Yukawa coupling can cause a similar enhancement of
BR($\tau\to lll)$. If a signal were established
for $\tau\to lll$ the angular distribution
of the leptons can act as powerful discriminator of the models
\cite{Kitano:2000fg}.
Such studies could be carried out at a high luminosity
$e^+e^-$ B factory \cite{Hashimoto:2004sm}. 

Our work is organized as follows. In section 2 the
manifest LR symmetric model is briefly reviewed. In section 3 
a numerical analysis of BR$(\tau\to lll)$ and BR($\mu\to e\gamma$) is
presented. In section 4 we discuss how to discriminate
the LR model from other models which contain a $H^{\pm\pm}$ 
by means of angular asymmetries in the LFV decays.
Conclusions are contained in section 5.

%%%%%%%%%%%%%%%%%%%%%%%%%%%%%%%%%%%%%%%%%
%
\section{Left-Right Symmetric Model}
%
%
%%%%%%%%%%%%%%%%%%%%%%%%%%%%%%%%%%%%%%%%%

%The smallness of the neutrino masses (with respect to the
%charged fermions) is elegantly and naturally explained
%by the celebrated seesaw mechanism in which an extremely heavy 
%Majorana mass ($>10^8$ GeV) for the right-handed
%neutrino together with an electroweak scale Dirac mass term leads
%to very light masses for the observed neutrinos.
%However, such heavy right-handed neutrinos are beyond the reach of 
%the LHC. Various alternative mechanisms have been proposed for neutrino mass
%generation at energies accessible to the LHC (see Smirnov 
%hep-ph/0411194 for a succinct review)

The Left-Right (LR) symmetric model
is an extension of the Standard Model (SM)
based on the gauge group $SU(2)_R \otimes SU(2)_L \otimes U(1)_{B-L}$.
The LR symmetric model has many virtues, e.g. i) the restoration of 
parity as an original symmetry of the Lagrangian which is broken 
spontaneously by a Higgs vev, and ii) the replacement of the
arbitrary SM hypercharge $Y$ by the theoretically more attractive 
$B-L$. Although the Higgs sector is
arbitrary, a theoretically and phenomenologically 
appealing way to break the $SU(2)_R$ gauge symmetry 
is by invoking Higgs isospin triplet representations. 
Such a choice conveniently 
allows the implementation of a low energy seesaw mechanism
for neutrino masses. The vev of the neutral member of the
right-handed triplet ($v_R$) can be chosen to give a TeV scale
Majorana mass term for the right-handed neutrinos, 
while the bidoublet Higgs fields provide the
small Dirac mass, leading to light masses for the observed neutrinos.
The above LR symmetric model predicts several new particles, 
among which the new gauge bosons $W_R$, $Z_R$ and doubly charged scalars
$H^{\pm\pm}_L$, $H^{\pm\pm}_R$ \cite{Gunion:1996pq}
have impressive discovery potential at hadron colliders if $v_R={\cal O}$ 
(1-10) TeV due to their large cross-sections 
and/or low background signatures.

Experiments which search for LFV
decays of $\mu$ and $\tau$ provide a complementary way of probing
the LR symmetric model. A comprehensive
study of $\mu\to e\gamma$, $\mu\to eee$ and $\mu\to e $ conversion
in the present model was performed in \cite{Cirigliano:2004mv}.
However, the recent
termination of the MECO ($\mu\to e$ conversion) experiment
together with no immediate improvement for the SINDRUM 
collaboration limit BR($\mu\to eee)<10^{-12}$
\cite{Bellgardt:1987du} leaves 
$\mu\to e\gamma$ as the only means of testing the
LR model in LFV processes involving $\mu$ in the near future
\cite{Grassi:2005ac}.

An alternative probe of the LR model which was not developed
in \cite{Cirigliano:2004mv}
are the LFV decays $\tau\to l_il_jl_k$. Although
the experimental sensitivity is inferior to
that for the above processes involving $\mu$, 
the decays $\tau\to l_il_jl_k$
have the virtue of probing many combinations of the triplet 
Higgs-lepton-lepton Yukawa $h_{ij}$ couplings. 
We introduce various strutures for the arbitrary $h_{ij}$ motivated by
the currently preferred bi-large mixing form of the 
Maki-Nakagawa-Sakata \cite{Maki:1962mu} (MNS) matrix.  
Should a signal for $\tau\to lll$ and/or $\mu\to e\gamma$ be observed,
the angular distribution of the final state leptons can provide 
a means of distinguishing models with $H^{\pm\pm}$.
We now briefly introduce the LR model and present the relevant
formulae for the numerical discussion in Section 3.
For a detailed introduction we refer the reader to
\cite{Duka:1999uc}.

The quarks and leptons are assigned to multiplets with quantum numbers 
($T_L, T_R, B-L$):
\beqa
\label{eq:matter}
 Q_{iL}&=&\left( \begin{array}{c}
u^\prime_i\\d^\prime_i\end{array}\right)_{L}: \left(1/2:0:1/3\right),~
Q_{iR}=\left( \begin{array}{c}
u^\prime_i\\d^\prime_i\end{array}\right)_{R}: \left(0:1/2:1/3\right) \ , \nn \\ 
L_{iL}&=&\left( \begin{array}{c}
\nu^\prime_i\\l^\prime_i\end{array}\right)_{L}: \left(1/2:0:-1\right),~
L_{iR}=\left( \begin{array}{c}
\nu^\prime_i\\l^\prime_i\end{array}\right)_{R}: \left(0:1/2:-1\right) \,.
\eeqa
Here $i=1,2,3$ denote generation number.
The spontaneous symmetry breaking to $U(1)_{em}$ occurs
through the Higgs mechanism. The Higgs sector consists of
a bidoublet Higgs field, $\Phi$,
and two triplet Higgs fields, $\Delta_L$ and $\Delta_R$:
\beqa 
\Phi=\left( \begin{array}{cc} \phi_1^0 & \phi_2^+\\\phi_1^- &
\phi_2^0\end{array}\right)&:& \left(1/2:1/2:0\right) \ , \nn \\
\Delta_{L}=\left( \begin{array}{cc}
\delta_{L}^+/\sqrt{2} & \delta_{L}^{++} \\ \delta_{L}^0 &
-\delta_{L}{^+}/\sqrt{2}
\end{array}\right): \left(1:0:2\right)  \ ,
\Delta_{R}&=&\left( \begin{array}{cc}
\delta_{R}^+/\sqrt{2} & \delta_{R}^{++} \\ \delta_{R}^0 &
-\delta_{R}{^+}/\sqrt{2}
\end{array}\right): \left(0:1:2\right) \ . 
\eeqa
%%%%%%%%%%
The vevs for these fields are as follows:
\beqa 
\langle\Phi\rangle&=&\left( \begin{array}{cc} \kappa_1/\sqrt{2} & 0\\ 0 &
\kappa_2/\sqrt{2}\end{array}\right),~~ 
\langle\Delta_{L}\rangle=\left( \begin{array}{cc}0& 0\\ v_{L}/\sqrt{2} & 0\end{array}\right) \ ,~~ 
\langle\Delta_{R}\rangle=\left( \begin{array}{cc}0& 0\\ v_{R}/\sqrt{2} & 0\end{array}\right) \ .
\eeqa
The gauge groups 
$SU(2)_R$ and $U(1)_{B-L}$ are spontaneously broken at the scale $v_R$.
Phenomenological considerations require $v_R \gg \kappa 
=\sqrt{\kappa_1^2+\kappa_2^2}\sim \frac{2M_{W_1}}{g}$ (EW scale).
The vev $v_L$ does not play a role in the breaking of the 
gauge symmetries and is constrained to be small ($v_L < 8$ GeV)
in order to comply with the measurement of 
$\rho=M_Z\cos\theta_W/M_W\sim 1$. 
The LR model predicts six neutral Higgs bosons, 
two singly charged Higgs bosons, and two
doubly charged Higgs bosons.
%$H_L^{\pm\pm}, H_R^{\pm\pm}, H_1^{\pm}, H_2^{\pm}, G_1, G_2 $
The Lagrangian is required to be the invariant under the 
discrete left-right symmetry:
$Q_L \leftrightarrow Q_R$\ ,  $L_L \leftrightarrow L_R$\ ,
$\Delta_L \leftrightarrow \Delta_R$\ , $\Phi \leftrightarrow \Phi^\dag$. 
This ensures equal gauge couplings ($g_L=g_R=g$) for 
$SU(2)_L$ and $SU(2)_R$.

The leptonic Yukawa interactions are as follows:
%
%%%%%%%%%%%%%%%%%%%%%%%%%%%%%%%%%%%%%%%%%%
\beq
-{\cal L}^{yuk}=\bar L_L(y_D\Phi+\tilde y_D\tilde \Phi)L_R 
+iy_M(L_L^TC\tau_2\Delta_LL_L+L_R^TC\tau_2\Delta_RL_R)+h.c.\ .
\eeq
Here $\tilde \Phi\equiv\tau_2\Phi^\ast\tau_2$; $y_D,\tilde y_D$ are
Dirac type Yukawa coupling;
$y_M$ is a $3\times 3$ Majorana type Yukawa coupling matrix
which will lead to Majorana neutrino masses (see below) and is 
the primary motivation for introducing the Higgs triplet 
representations $\Delta_L$ and $\Delta_R$. 
Invariance under the left-right discrete symmetry gives 
$y_D=y_D^\dag$, $\tilde y_D=\tilde y_D^\dag$ and $y_M=y_M^T$. 
Redefinitions of the fields $L_L$ and $L_R$ enable $y_M$ to be taken as real,
positive and diagonal, while maintaining 
$y_D=y_D^\dag$, $\tilde y_D=\tilde y_D^\dag$. Hereafter $y_M$ is taken in this
diagonal basis.
The $3\times 3$ mass matrix for charged leptons is:
\beqa
M_l=\frac{1}{\sqrt{2}} (y_D\kappa_2+\tilde y_D\kappa_1)\ ,
\eeqa
which is diagonalized by the unitary matrices, $V_L^l$ and $V_R^l$, as:
\beq
V_L^{l\dag} M_lV_R^l=diag(m_e, m_\mu, m_\tau)\ .
\eeq
The Lagrangian for the neutrino masses is: 
\beq
-{\cal L}_{mass}=\frac{1}{2}(\bar n_L M_\nu n_R+\bar n_RM_\nu^\ast n_L) \,,
\eeq
where $n_L=(\nu_L,\nu_R^c)^T$ and $n_R=(\nu_L^c, \nu_R)^T$ with the definition of
$\nu_R^c %=(\nu_R)^c
=C(\bar {\nu_R})^T$.
The $6 \times 6$ mass matrix for the neutrinos can be written 
in the block form:
\beq
M_\nu=\left( \begin{array}{cc}
M_L & m_D
\\
m_D^T & M_R
\end{array}\right).
\label{eq:Mnu}
\eeq
Each entry is given by:
\beqa
m_D
&=&{1\over \sqrt{2}}\left(y_D\kappa_1+{\tilde y_D \kappa_2}\right)\ , \\
M_R&=&\sqrt{2}y_M v_R \ , \\
M_L&=&\sqrt{2}y_M v_L \ . 
\eeqa
The neutrino mass matrix is diagonalized by a $6\times 6$ unitary matrix 
$V$
as $V^TM_\nu V=M_\nu^{diag}=diag(m_1,m_2,m_3,M_1,M_2,M_3)$, where $m_i$ and $M_i$
are the masses for neutrino mass eigenstates:
\beqa
V\equiv
\bmaT
V_{L}^{\nu\ast} &V_{L}^{\nu'\ast} \\
V_{R}^{\nu'} &V_{R}^{\nu} 
\emaT\ .
\eeqa
The small neutrino masses $m_i$ are generated by the 
Type II seesaw mechanism. Obtaining eV scale neutrino masses 
with $y_M={\cal O} (0.1-1)$
requires $M_L$ (and consequently $v_L$)
to be eV scale. However, the minimization of the Higgs potential 
% $\beta_i\Phi\Delta_R\Phi^T\Delta_L$
leads to a relationship among the vevs, $v_L\sim \gamma\kappa^2/v_R$, 
where $\gamma$ is a function (introduced in \cite{Deshpande:1990ip})
of scalar quartic couplings $\beta_i$ and $\rho_i$.
%$\begin{equation}
%\frac{\beta_2\kappa_1^2+\beta_1\kappa_1\kappa_2+\beta_3\kappa_2^2}
%{(2\rho_1-\rho_3)\kappa^2
%\end{equation}
For natural values of $\beta_i$ and $\rho_i$ one has 
$\gamma\sim 1$ and thus $v_L$ would be ${\cal O} (1-10)$ GeV for 
$v_R\sim$ TeV. Reducing $v_L$ to the eV scale to order to
comply with the observed neutrino mass scale would
require severe fine-tuning  $\gamma < 10^{-7}$.
In LR model phenomenology it is standard to set $\beta_i=0$
(and hence $\gamma=0$) which ensures $v_L=0$.
Henceforth we will take $v_L=0$ for which
the masses of the light neutrinos arise from Type I seesaw mechanism and are approximately
$m_i\sim m_D^2/M_R$. 
\footnote{For an alternative approach which maintains the Type II
seesaw mechanism and obtains 
$v_L\sim $eV by means of horizontal symmetries
see \cite{Kiers:2005vx}.}
In order to realize the low energy 
($\sim {\cal O}(1-10)$ TeV) scale for the right-handed
Majorana neutrinos, 
the Dirac mass term $m_D$ should be {$\cal O$} (MeV), 
which for $\kappa_2\sim 0$
corresponds to $y_D\sim 10^{-6}$ 
(i.e. comparable in magnitude to
the electron Yukawa coupling). 

There are two physical singly charged Higgs bosons,
$H^\pm_1$ and $H^\pm_2$, which are linear combinations of
the singly charged scalar fields residing in 
$\Phi,\Delta_L$ and $\Delta_R$.
The leptonic couplings $\tilde y_D$ of $H_2^\pm$ (which
is essentially composed of $\phi_1^\pm$ and $\phi_2^\pm$)
are of order $m_l/m_W$ and
can be neglected compared to leptonic Yukawa couplings
for the triplet field $H^\pm_1\sim \delta_L^\pm$
which are unrelated to fermion masses and may be
sizeable. The interaction of $H_1^\pm$ with leptons is 
as follows (where  $N_L=V^Tn_L$, $N_R=V^\dagger n_R$, 
$N=N_L+N_R=N^c$ and $l=l_R+l_L$ are the neutrino
and charged lepton fields respectively in the mass eigenstate basis,
and $P_{L,R}=(1\mp \gamma_5)/2$):
\beqa
{\cal L}_{H_1^\pm
%\delta_L^{\pm}
} &=& %\displaystyle\frac{g}{
\sqrt{2} %} \ 
\Bigg[
H_{1}^+  \
\overline{N}  \left( \tilde{h} \, P_L
\right) l
+ H_1^-  \
\overline{l}  \left( \tilde{h}^\dagger  \, P_R
\right) N
\Bigg]\ .
\label{eq:singly}  
\eeqa
The LFV interactions of leptons with doubly 
charged Higgs bosons (where $H^{\pm\pm}_L=\delta^{\pm\pm}_L,
H^{\pm\pm}_R=\delta^{\pm\pm}_R$ for $v_L=0$) are given by:
\beqa
{\cal L}_{H^{\pm \pm}_{L,R}} &=&
%\displaystyle\frac{g}{2} \ 
\Bigg[
H_{L,R}^{++}  \
\overline{l^{c}}  \left( h_{L,R} \, P_{L,R} \right) l      +
H_{L,R}^{--}  \
\overline{l}  \left( h_{L,R}^{\dagger} \, P_{R,L} \right) l^{c}
\Bigg]
\label{eq:doubly} \  .
\eeqa
 The LFV coupling matrices in Eq.(\ref{eq:singly}) and Eq.(\ref{eq:doubly}) 
are respectively given by: 
\beqa
\tilde h&=&\frac{1}{\sqrt{2}v_R} \left( \begin{array}{c} V_L^{\nu T} \\ V_L^{\nu' T} \end{array}\right)
M_RV_L^l \,,    \\
h_L&=&\frac{1}{\sqrt{2}v_R}
V_L^{l T}M_RV_L^l  \,, 
\label{eq:hL}\\
h_R&=&\frac{1}{\sqrt{2}v_R}
V_R^{l T}M_RV_R^l \,.
%=\frac{V_R^{l T}M_RV_R^l}{\sqrt{2}v_R}
%\sim V_R^{l T}V_R^{\nu \ast} \frac{M}{\sqrt{2}v_R}V_R^{\nu\dag}V_R^l \,.
\label{eq:hR}
\eeqa
Note that $\tilde h$ is a $6\times 3$ matrix and $h_L$ and $h_R$ are
$3\times 3$ matrices.
 
%Here $V_R^{\nu T}M_RV_R^\nu\simeq M=diag(M_1,M_2,M_3)$.
The mass matrix for the charged vector bosons is:
\beq
{\tilde M}_W^2={g^2\over 4}\left( \begin{array}{cc}
\kappa^2 & -2\kappa_1\kappa_2
\\
-2\kappa_1\kappa_2 & \kappa^2+2v_R^2
\end{array}\right) ~.
\eeq
This is diagonalized via the mixing angle
$\xi=-\tan^{-1}\left(2\kappa_1\kappa_2/v_R^2\right)/2$ with the
eigenvalues $M_{W_{1,2}}^2=g^2\left(\kappa^2+v_R^2 \mp
\sqrt{v_R^4+4\kappa_1^2\kappa_2^2}\right)/4$:
\beq
W_L = \cos \xi \, W_1  + \sin \xi \, W_2, \qquad
W_R = - \sin \xi \, W_1  + \cos \xi \, W_2 \ . 
\eeq 
The strong experimental constraint on the
mixing angle ($\xi< 10^{-3}$)  \cite{Yao:2006px}
enforces one of $\kappa_1,\kappa_2$ to be small if
$v_R={\cal O}$ (TeV).
Neglecting such small mixing between $W_1$ and $W_2$,
the LFV interactions with the gauge bosons are as follows: 
\beqa
{\cal L}_{CC} &=& \displaystyle\frac{g}{\sqrt{2}} \Bigg\{
\overline{N} \Big[ \gamma^\mu \, P_R \, (K_R) \Big] l  \cdot
W_{2 \ \mu}^+
+
\overline{l} \Big[ \gamma^\mu \, P_R \, (K_R^\dagger) \Big] N \cdot
W_{2 \ \mu}^-
%\Bigg\}  
\nn \\
&&+\overline{N} \Big[ \gamma^\mu \, P_L \, (K_L) \Big] l  \cdot
W_{1 \ \mu}^+
+
\overline{l} \Big[ \gamma^\mu \, P_L \, (K_L^\dagger) \Big] N \cdot
W_{1 \ \mu}^-
\Bigg\}  \ ,
\eeqa
where
$K_L$ and $K_R$ are the $6\times 3$ LFV coupling
matrices which are respectively written as:
\beq
K_L= \bmaT\begin{array}{c} V_L^{\nu\dag} V_L^l \\  V_L^{\nu'\dag} V_L^l \end{array}\emaT 
\simeq\bmaT\begin{array}{c} V_{\rm MNS}^\dag \\  V_L^{\nu'\dag} V_L^l \end{array}\emaT \ ,
~~~
K_R=\bmaT\begin{array}{c} V_R^{\nu '\dag} V_R^l \\  V_R^{\nu \dag} V_R^l \end{array}\emaT \ .
\label{eq:K}
\eeq
The upper $3\times 3$ block in $K_L$ can be identified 
as the hermitian conjugate of
the MNS matrix $V_{\rm MNS}$ on neglecting ${\cal O}(\frac{m_D}{M_R})$ contributions.

%%%%%%%%%%%%%%%%%%%%%%%%%%%%%%%
%
\subsection{Manifest LR Symmetric Model}
%
%%%%%%%%%%%%%%%%%%%%%%%%%%%%%%%
In the LR model the mixing matrices for the
left and right fermions are in general not equal 
e.g. for the lepton sector $V_L^l\ne V_R^l$.
The special case of $V_L^l= V_R^l$ is referred to
as the ``Manifest LR symmetric model'' and arises 
in either of the following scenarios: 
i) both $\kappa_1$ and $\kappa_2$
are real, or 
ii) one of $\kappa_1$ and $\kappa_2$ is identically zero.
In our numerical analysis we will set $\kappa_2=0$, which 
has the virtue of eliminating $W_L-W_R$ mixing and 
in some cases (for specific forms of the Higgs potential) 
is required to suppress FCNCs and preserve unitarity in the LR model
\cite{Gunion:1989in}.
In the Manifest LR symmetric model one has the additional
constraint $m_D=m_D^\dag$ which must be respected when
evaluating the magnitude of the LFV processes.
%However, such inequalities 
%are a consequence of CP violation and arise only if 
%there is a relative phase between $\kappa_1$ and $\kappa_2$. 
%In the LR model one may take $\kappa_2=0$, which 
%has the virtue of eliminating $W_L-W_R$ mixing and 
%in some cases (for specific forms of the Higgs potential) 
%is required to suppress FCNCs and preserve unitarity in the LR model.
%In such a scenario the relative phase between $\kappa_1$ 
%and $\kappa_2$ can be removed, leading to the 
%``Manifest LR symmetric model'' in which $V_L^l= V_R^l$ 
%and $m_D=m_D^\dag$. Henceforth we shall
%take $\kappa_2=0$.
% for which the additonal constraint 
%$m_D=m_D^\dag$ plays a key role in our numerical analysis.

A further important consequence of the Manifest LR symmetric model
is the relationship $h_L=h_R\equiv h$ which can be derived
from Eq.(\ref{eq:hL}) and Eq.(\ref{eq:hR}).
Using $K_L$ and $K_R$ in Eq.(\ref{eq:K}), 
the LFV couplings for the interactions of leptons
with singly and doubly charged Higgs bosons are as follows: 
\beq
\tilde{h} = K_L^*h\ , \,   ~~~~~
h =\frac{1}{\sqrt{2}v_R} K_R^T \, M_{\nu}^{\rm diag} \, K_R  \ .
\label{eq:hcouplings}
\eeq
At leading order in $m_D/M_R$, one may express $h$ by:
\beq 
h_{i j} = \sum_{n={\rm heavy}} \, \Big( K_R \Big)_{n i}  
\Big( K_R \Big)_{n j}  \, \sqrt{x_n}  \ , 
\label{eq:hcouplings1}
\eeq
\beq
 x_n = \left({M_n\over \sqrt{2}v_R}\right)^2\  .
\eeq

%
%%%%%%%%%%%%%%%%%%%%%%%%%%%%%%%%%%%%%%%%%%
%
%      LFV decays
%
%%%%%%%%%%%%%%%%%%%%%%%%%%%%%%%%%%%%%%%
\subsection{Effective Lagrangian and branching ratios for the LFV processes}

%\bea
%{\cal L}=-\frac{4G}{\sqrt{2}}\left\{m_\tau A_R\ov\tau\sigma^{\mu\nu}P_L\mu F_{\mu\nu}
%+m_\tau A_L\ov\tau \sigma^{\mu\nu}P_R\mu F_{\mu\nu}\right. \nn \\
%\left.
%+g_3(\ov\tau\gamma^\mu P_R\mu)(\ov\gamma_\mu P_R\mu) 
%+g_4(\ov\tau\gamma^\mu P_L\mu)(\ov\gamma_\mu P_L\mu)+h.c. 
%\right\}
%\eea
%%%%%%%%%  mu or tau -> lll %%%%%%%%%%%%%%%%
%
\subsubsection{4-lepton interactions}
%$\tau^\pm \to l_1^\pm l_2^\pm l_3^\mp$ and $\mu \to 3e$}
%
The effective Lagrangian for %$\tau\to lll$ 
4-lepton interactions is as follows:
\beqa
{\cal L}\hspace{-3mm}
%&=&\hspace{-3mm}\left(H^{++}_L\overline{(l_L)^c}h_Ll_L\right)H^{--}\overline l(h_L)^\dag
%(l_L)^c+(L\leftrightarrow R) \nn \\
&=&\hspace{-3mm}
\frac{1}{2}(h^\ast)_{mi}(h)_{jk}
\left\{
\frac{1}{M_{H^{\pm\pm}_L}^2}
\left( \overline l_m\gamma^\mu P_Ll_{k} \right)\left( \overline l_i\gamma_\mu P_Ll_{j} \right)
+\frac{1}{M_{H^{\pm\pm}_R}^2}
\left( \overline l_m\gamma^\mu P_Rl_{k} \right)\left( \overline l_i\gamma_\mu P_Rl_{j} \right) 
\right\}
\ .
\eeqa
%Then, 
%\bea
%g_3\equiv hh^\dag
% \eea

The branching ratio for $\tau \to l_il_jl_k$ is given by:
\beq
BR(\tau \to l_il_jl_k)=\frac{8S}{g^4}|h^\ast_{\tau l_i}h_{l_jl_k}|^2
\left(\frac{M^4_{W_1}}{M^4_{H^{\pm\pm}_L}}+\frac{M^4_{W_1}}{M^4_{H^{\pm\pm}_R}}
 \right)BR(\tau\to \mu\nu\nu)\ .
\label{BRtaulll}
\eeq
Here $S$=1 (2)  for $j=k$ ($j\ne k$). 
In the LR model one may express $h_{ij}$ in terms of $K_RK_R$
via Eq.(\ref{eq:hcouplings1}).
However, the above form Eq.(\ref{BRtaulll}) 
can be applied to other models with $H_L^{\pm\pm}$ and $H_R^{\pm\pm}$
for which no identity Eq.(\ref{eq:hcouplings1}) exists.

%%%%%%%%%%%%%%  mu -> e \gamma   %%%%%%%%%%%%%
\subsubsection{$\mu \to e\gamma$}

The effective Lagrangian for $\mu \to e \gamma$ is as follows:  
\beq
{\cal L}=-\frac{4G}{\sqrt{2}}\left\{m_\mu A_R\overline\mu\sigma^{\mu\nu}P_Le F_{\mu\nu}
+m_\mu A_L\overline\mu \sigma^{\mu\nu}P_Re F_{\mu\nu}+h.c. 
\right\} \ .
\eeq
$A_L$ receives contributions from $W_2-N_i$ and $H^{\pm\pm}_R$ and
is given by \cite{Cirigliano:2004mv}:
\beqa
A_L & = & \frac{1}{16 \pi^2} \, 
\displaystyle\sum_{n={\rm heavy}} \, \Big( K_R^\dagger \Big)_{e n} 
\Big( K_R \Big)_{n \mu}  \ 
\left[ \frac{M_{W_1}^2}{M_{W_2}^2}  \, S_3 (x_n) 
-  \frac{x_n}{3}  \frac{M_{W_1}^2}{M_{H_R^{\pm\pm}}^2}  
\right] \ , 
\label{mu_ey_AL}
\eeqa
where
\beq
 S_3 (x)=-\frac{x(1+2x)}{8(1-x^2)^2}+\frac{3x^2}{4(1-x)^2}\left\{\frac{x}{(1-x)^2}(1-x+\log x)+1\  \right\}\ .
\eeq
$A_R$ receives contributions from $H^{\pm\pm}_L$ and $H_1^\pm$
and is given explicitly by \cite{Cirigliano:2004mv}:
\beqa
A_R & = & \frac{1}{16 \pi^2} \, 
\displaystyle\sum_{n={\rm heavy}} \, \Big( K_R^\dagger \Big)_{e n} 
\Big( K_R \Big)_{n \mu}  \ \, x_n 
\left[  -  \frac{1}{3}  \frac{M_{W_1}^2}{M_{H_L^{\pm\pm}}^2}  
-  \frac{1}{24}  \frac{M_{W_1}^2}{M_{H_1^\pm}^2}  
\right] \ . 
\label{mu_ey_AR}
\eeqa
The branching ratio for $\mu \to e\gamma$ is given 
by: 
\footnote{In our numerical analysis we do not  
include a suppression factor of
$\sim 15\%$ arising from electromagnetic
corrections \cite{Czarnecki:2001vf}.}
\beq
BR(\mu\to e\gamma) =384\pi^2e^2(|A_L|^2+|A_R|^2) .
\eeq

%%%%%%%%%%%%%%%%%%%%%%%%%%%%%%%%%%%%%%%%%%%%%%

%%%%%%%%%%%%%%%%%%%%%%%%%%%%%%%%%%%%%%%%%%%%%%
%
%          Numerical analysis for the lepton flavor violation
%         
%
%%%%%%%%%%%%%%%%%%%%%%%%%%%%%%%%%%%%%%%%%%%%%%

\section{Numerical analysis for BR($\tau\to lll)$ and BR($\mu\to e\gamma$)}

The LFV decays $\tau\to lll$ are the analogy of $\mu\to eee$
and provide sensitive probes of the $h_{ij}$ couplings in the LR model.
Mere observation of such a decay would constitute a spectacular
signal of physics beyond the SM. 
There are six distinct decays for $\tau^+\to lll$ (likewise for $\tau^-$):
$\tau^+\to \mu^+\mu^+\mu^-$, $\tau^+\to e^+e^+e^-$,
$\tau^+\to \mu^+\mu^+e^-$, $\tau^+\to \mu^+\mu^-e^+$,
$\tau^+\to e^+e^+\mu^-$, $\tau^+\to e^+e^-\mu^+$.
Searches for all six decays have been performed by BABAR (91 fb$^{-1}$)
\cite{Aubert:2003pc}
and BELLE (87 fb$^{-1}$) \cite{Yusa:2004gm}.
%One candidate event for $\tau^-\to e^+e^-e^-$
%was observed by BELLE, with zero candidate events 
%for the remaining 5 decay modes. BABAR observed one candidate event 
%for each of $\tau^-\to e^+e^-e^-$, $\tau \to \mu^-\mu^+e^-$ and
%$\tau^-\to e^-e^+\mu^-$, with zero events for the remaining
%3 decay modes. 
Upper limits of the order BR($\tau\to lll) < 2\times10^{-7}$ were derived.
Although these limits are several orders of magnitude weaker 
than the
bound BR$(\mu\to eee)< 10^{-12}$, they have the virtue of
constraining many combinations of the $h_{ij}$ couplings in the 
context of the LR model. Moreover, 
greater sensitivity to BR($\tau\to lll$) is 
expected from forthcoming experiments.
A proposed Super B Factory anticipates sensitivity
to BR($\tau\to lll)\sim 10^{-8}$ and $10^{-9}$ for 5 ab$^{-1}$
and 50 ab$^{-1}$ respectively \cite{Hashimoto:2004sm}. 
At the LHC, $\tau$ can be copiously produced from several sources
(from $B/D$ decay and direct production via $pp\to W\to \tau\nu$,
$pp\to Z\to \tau^+\tau^-$)
and sensitivity to BR$(\tau\to lll)>10^{-8}$ is claimed 
\cite{Santinelli:2002ea}.
Such low BRs can be reached due to the very small SM background.  
In contrast, the background 
to $\tau\to l\gamma$ is non-negligible and might prevent
the B factories from probing below BR$\sim 10^{-8}$.
In addition, one expects BR$(\tau\to lll) \gg {\rm BR}(\tau\to l\gamma)$ in the
LR symmetric model and thus the former decay is the more effective probe.
We note that other rare LFV $\tau$ decays involving quark final states will
not arise in the LR model since $h_{ij}$ mediates processes involving
leptons only. Hence we shall only focus on $\tau\to lll$.
In our numerical analysis only the stringent constraint from 
$\mu\to eee$ is imposed. Other constraints on $h_{ij}$ 
(e.g. the anomalous magnetic moment of $\mu$ ($g-2$), Bhabha scattering and
other LFV processes - see \cite{Cuypers:1996ia})
are considerably weaker and are neglected.

The magnitude of $h_{ij}$ cannot be predicted from the 
neutrino oscillation data alone since it is related to the physics 
at SU(2$)_R$ breaking scale. However, $h_{ij}$ also 
crucially depends on the
mixing matrix in the charged lepton sector $V^l_L$:
\beq
h=\frac{1}{\sqrt{2}v_R}V_L^{lT}M_RV_L^l \ .
\label{h_coup}
\eeq
We have neglected the ${\cal O}(\frac{m_D}{M_R})$ contribution and used the convention 
that $M_R$ is diagonal and positive.
%where $M_R$ is diagonal and positive \cite{Duka:1999uc},\cite{Kiers:2005vx}. 
Since $V_L$ also enters the MNS matrix:
\beq
V_{\rm MNS}=V_L^{l\dag}V_L^\nu \ ,
\label{MNS:V_L}
\eeq
we will introduce 4 distinct structures for $V^l_L$ motivated by the bi-large 
mixing form of $V_{\rm MNS}$ and perform a quantitative analysis of the
magnitude of $h_{ij}$ (and consequently BR($\tau\to lll$)) in the LR model.

In order to establish our formalism we first explicitly present the 
standard parametrisation of the MNS matrix:
\beqa
V_{\rm MNS}&=&
\left(\begin{array}{ccc}1&0&0\\0&c_{23}&s_{23} \\0&-s_{23}&c_{23}  \\\end{array}\right)
\left(\begin{array}{ccc}c_{13}&0&s_{13}e^{-i\delta}\\0&1&0 \\-s_{13}e^{i\delta}&0&c_{13}  \\\end{array}\right)
\left(\begin{array}{ccc}c_{12}&s_{12}&0 \\-s_{12}&c_{12}  &0\\0&0&1\end{array}\right)
\nn \\
&=&U(\theta_{23})U(\theta_{13})U(\theta_{12})\ .
\label{eq:MNS}
\eeqa
Here $s_{ij}$ $(c_{ij})$ represents $\sin\theta_{ij}$ $(\cos\theta_{ij})$
and the unitary matrices
$U(\theta_{23})$, $U(\theta_{13})$, and $U(\theta_{12})$ are 
responsible for mixing between 2-3, 1-3, and 1-2 elements respectively. 
%(the mixing angle of $\theta_{23}$, $\theta_{13}$, and $\theta_{12}$. 

The angles $\theta_{12}$ and $\theta_{23}$ are 
measured with relatively good accuracy in the solar and atmospheric neutrino 
oscillation experiments respectively. 
The solar and KamLAND reactor neutrino oscillation experiments 
\cite{Aharmim:2005gt},\cite{Araki:2004mb} provide the 
following constraints on the mixing angle 
$\theta_{12}$ and the mass-squared difference of $\Delta m_{12}^2=m_2^2-m_1^2$:
%\beq
$
\sin^2\theta_{12}\sim 0.31\ , ~\Delta m_{12}^2\sim 8\times 10^{-5} ~{\rm eV}^2\ . 
$
%\eeq
The mixing angle $\theta_{23}$ and the mass-squared difference 
$\Delta m_{13}^2$
measured in the atmospheric neutrino oscillation are as follows
\cite{Ashie:2005ik},\cite{Aliu:2004sq}:
%\beq
$
\sin^22\theta_{23}\sim 1.0\ , ~|\Delta m_{13}^2|\sim 2.6\times 10^{-3}~{\rm eV}^2\ .  
$
%\eeq
An upper bound on
the remaining angle $\theta_{13}$ has been obtained from the 
CHOOZ and Palo Verde reactor neutrino oscillation experiments 
\cite{Apollonio:2002gd},\cite{Boehm:2001ik}:
%\beq
$
\sin\theta_{13} \lsim 0.2\ .
$
%\eeq
%The T2K neutrino oscillation experiment anticipates sensitivity to $\sin^22\theta_{13}> 0.006$.

The ignorance of the sign of $\Delta m_{13}^2$ and the 
absolute neutrino mass scale leads to the following three neutrino 
mass patterns 
which are consistent with current oscillation data: 
Normal hierarchy (NH)  $m_1 < m_2 \ll m_3$; 
Inverted hierarchy (IH)  $m_3 \ll m_1 < m_2$; 
Quasi degeneracy (DG) $m_1 \sim m_2 \sim m_3$.
Data from WMAP \cite{Spergel:2006hy}
provides the following constraint on the sum of the 
light neutrino masses:
$\sum_{i=1,2,3} m_i < 2$ eV.
However, LFV processes in the LR model do not depend 
sensitively on the neutrino mass pattern.

In order to perform our numerical analysis of the
magnitude of BR$(\tau\to lll)$ and BR$(\mu\to e\gamma)$
we introduce the following four specific cases:
\begin{center}
\begin{tabular}{c|ll}
CASE & $~~~V_L^{l\dag}$ & $V_L^\nu $ \\
\hline
I  &  $-iV_{\rm MNS}$ & $iI$ \\
II  & $ -iU(\theta_{23})U(\theta_{13})$ & $iU(\theta_{12})$\\
III & $-iU(\theta_{23})$ & $iU(\theta_{13})U(\theta_{12})$ \\
IV & $-iI$ & $iV_{\rm MNS}$ 
\end{tabular}
\end{center}
Here $I$ represents a unit matrix.
In CASE I (IV) both large mixings in $V_{\rm MNS}$
originate from the charged lepton (neutrino)
mixing matrix. Each case has distinct ways of
satisfying the stringent bound BR($\mu\to eee)<10^{-12}$.
In our numerical analysis 
we will assume multi-TeV scale masses for $H^{\pm\pm}_L$ and
$H^{\pm\pm}_R$ which renders direct detection improbable
at the LHC. For $M_{H^{\pm\pm}_L}$,$M_{H^{\pm\pm}_R}<1 $ TeV 
the LHC has excellent discovery prospects in the channels
$H^{\pm\pm}\to l^\pm_i l^\pm_j$ \cite{Gunion:1996pq}
(especially for $l_{i,j}=e,\mu$). Observation
of $H^{\pm\pm}_{L,R}$ (which would provide a measurement of
$M_{H^{\pm\pm}}$) together with a signal for $\tau\to l_il_jl_k$
would permit a measurement of the coupling combination 
$|h_{\tau i}^\ast h_{jk}|$.

We wish to study the magnitude of BR($\tau\to lll$) and BR($\mu\to e\gamma$)
in the parameter space with a phenomenologically acceptable
neutrino mass matrix. In the LR model the light neutrino masses
arise from the seesaw mechanism and are approximately as follows: 
\beq
m_\nu=-m_DM_R^{-1}m_D^T\ .
\eeq
At the leading order, $m_\nu$ is diagonalized by $V_L^\nu$: 
\beqa
m_\nu&\simeq&V_L^\nu m_\nu^{diag}V_L^{\nu T}\ ,
\label{seesaw}
\eeqa
where $m_\nu^{diag}=diag(m_1,m_2,m_3)$. % and $m_{1,2,3} >0$.
The Dirac mass matrix $m_D$ depends on an arbitrary Yukawa coupling $y_D$.
As advocated in \cite{Casas:2001sr},
it is beneficial to parametrize a general seesaw type
matrix such that the arbitrary $m_D$ is replaced by potential
observables i.e. the heavy and light neutrino masses.
We will apply the formalism of \cite{Casas:2001sr} in
our numerical analysis, with the additional constraint
that the manifest LR model requires the Dirac mass matrix 
for both the neutrinos and charged leptons to be a hermitian matrix:
\beq
m_D=m_D^\dag \ .
\label{eq:mD_hermitian}
\eeq
We introduce a complex orthogonal matrix $R$ which satisfies $R^TR=1$ 
and parametrize the neutrino Dirac mass matrix $m_D$ as follows:
\beq
m_D=-iV_{L}^\nu\sqrt{m_\nu^{diag}}R^T\sqrt{M_R}\ .
\label{eq:mD}
\eeq
The LFV processes are evaluated in the parameter space where there
exists an $R$ matrix which satisfies the condition Eq.(\ref{eq:mD_hermitian}).
This condition guarantees a phenomenologically acceptable
neutrino mass matrix and perturbative Yukawa coupling $y_D$.
In our calculation, the Majorana phases in the MNS matrix are neglected while 
the CP conserving cases for the Dirac phase 
($\delta=0$ or $\pi$) are taken into account for simplicity.
We will comment on the case of the CP violating Dirac phase in section 3.5.
%Without loss of generality we work in a basis where the right-handed neutrino
% mass matrix $M_R$ 
%is a real diagonal matrix, $M_R=diag(M_1,M_2,M_3)$, 
%with $0 <M_1 \le M_2 \le M_3$.
Neglecting CP violation, the condition Eq.(\ref{eq:mD_hermitian}) requires
that $V_L^\nu$ is purely imaginary, 
while $R$ is a real matrix.
%According to this constraint we can avoid the unphysical parameter region
% where the Yukawa coupling for the neutrino Dirac mass grows up although such problem 
%occurs in the general cases. 
%{\it e.g} SUSY case.
We stress that the LFV processes do not depend on the actual
structure of $R$. However, proving the existence of an 
$R$ matrix for each of the four cases (I, II, III, IV) ensures 
the validity of our numerical analysis.
%We will assume multi-TeV scale masses for $H^{\pm\pm}_L$ and
%$H^{\pm\pm}_R$ which 

%%%%%%%%%%%%%%%%%%%%%%%%%%%%%%%%%%%%
\subsection{Numerical results: CASE I}
%%%%%%%%%%%%%%%%%%%%%%%%%%%%%%%%%%%%
The bi-large mixing originates from the charged lepton sector.
We parametrize the $R$ matrix as follows:
%Parametrizing the $R$ matrix as:
\beq
R=U(\theta^R_{23})U(\theta^R_{13})U(\theta^R_{12})\ .
\label{eq:Rmatrix}
\eeq
The explicit form in CASE I is given by:
\beq
\theta^R_{12}=\theta^R_{23}=\theta^R_{13}=0\ ,
\eeq
which leads to $R=I$. In this case $m_D=\sqrt{m_\nu^{diag}}\sqrt{M_R}$, 
and thus $m_D=m_D^\dag$ is automatically satisfied.
 
%%%%%%%%%%%%%%%%%%%%%%%%%%%%%%%%%%%%%%
%
%   Figure 1
%%%%%%%%%%%%%%%%%%%%%%%%%%%%%%%%%%%%%%
\begin{figure}[ht]
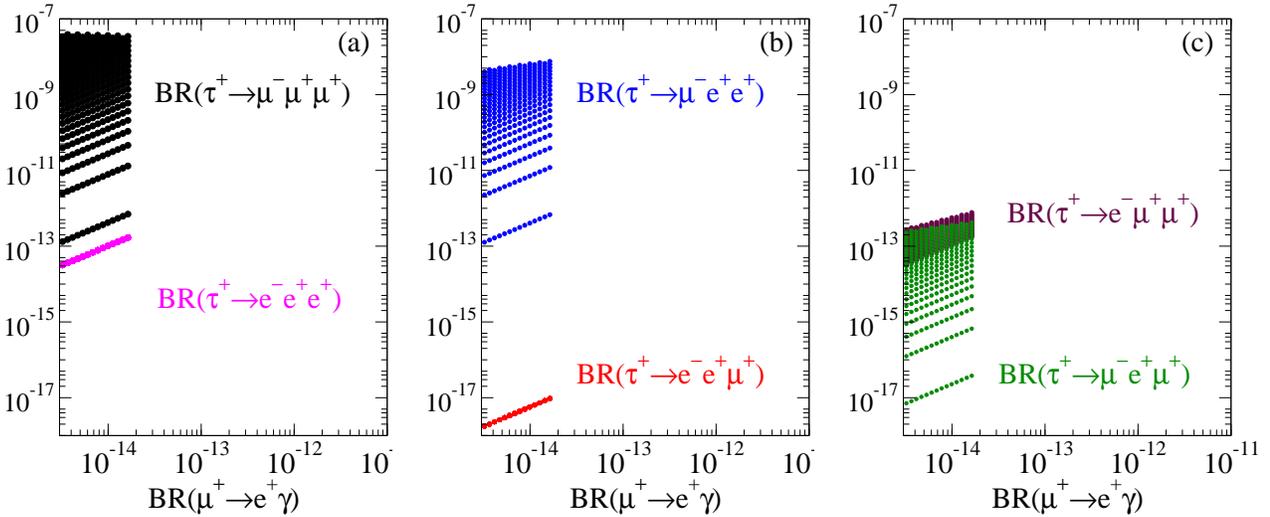

\centering
\begin{picture}(200,200)  
\put(70,75){\makebox(10,20){\epsfig{figure=case1_0a.eps,width=5.5cm}}}
\end{picture}
\hspace{-2.4cm}
\begin{picture}(200,200)  
\put(90,75){\makebox(10,20){\epsfig{figure=case1_0b.eps,width=5.5cm}}}
\end{picture}
\hspace{-2.4cm}
\begin{picture}(200,200)  
\put(110,75){\makebox(10,20){\epsfig{figure=case1_0c.eps,width=5.5cm}}}
\end{picture}
\caption{Branching ratios for 
(a) $\tau^+ \to e^-e^+e^+$ and $\tau^+ \to \mu^-\mu^+\mu^+$,  
(b) $\tau^+\to e^-e^+\mu^+$  and $\tau^+\to \mu^-e^+e^+$, and  
(c) $\tau^+\to e^-\mu^+\mu^+$  and $\tau^+\to \mu^-e^+\mu^+$ 
against BR$(\mu^+\to e^+\gamma)$ for $s_{13}=0$ in CASE I. The experimental bound on BR$(\mu\to eee)$ ($<10^{-12}$) is imposed. }
\label{fig:1}
\end{figure}

%%%%%%%%%%%%%%%%%%%%%%%%%%%%%%%%%%%%%
%
%    Figure 2 (s3 >0)
%%%%%%%%%%%%%%%%%%%%%%%%%%%%%%%%%%%%%
\begin{figure}[ht]
\centering
\begin{picture}(200,200)  
\put(70,75){\makebox(10,20){\epsfig{figure=case1_2_0a.eps,width=5.5cm}}}
\end{picture}
\hspace{-2.4cm}
\begin{picture}(200,200)  
\put(90,75){\makebox(10,20){\epsfig{figure=case1_2_0b.eps,width=5.5cm}}}
\end{picture}
\hspace{-2.4cm}
\begin{picture}(200,200)  
\put(110,75){\makebox(10,20){\epsfig{figure=case1_2_0c.eps,width=5.5cm}}}
\end{picture}
\caption{
Branching ratios for 
(a) $\tau^+ \to e^-e^+e^+$ and $\tau^+ \to \mu^-\mu^+\mu^+$,  
(b) $\tau^+\to e^-e^+\mu^+$  and $\tau^+\to \mu^-e^+e^+$, and  
(c) $\tau^+\to e^-\mu^+\mu^+$  and $\tau^+\to \mu^-e^+\mu^+$ 
against BR$(\mu^+\to e^+\gamma)$ for $\sin\theta_{13}=0.2$ with $\delta=0$ in CASE I.}
\label{fig:2}
\end{figure}
%%%%%%%%%%%%%%%%%%%%%%%%%%%%%%%%%%%%%
%
%    Figure 3  (s3 <0 )
%%%%%%%%%%%%%%%%%%%%%%%%%%%%%%%%%%%%%
\begin{figure}[ht]
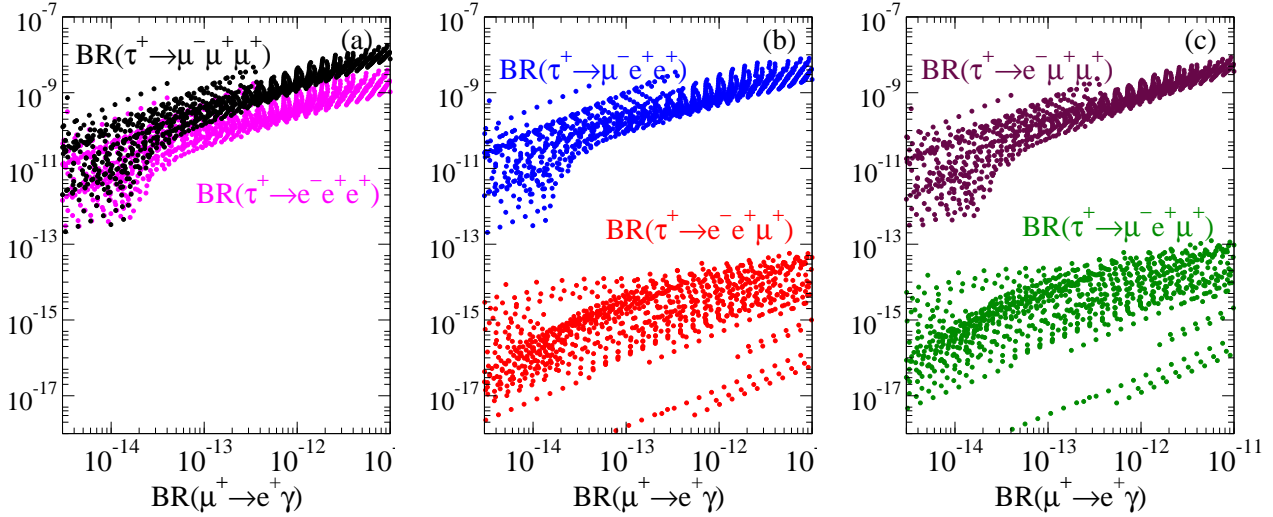

\centering
\begin{picture}(200,200)  
\put(70,75){\makebox(10,20){\epsfig{figure=case1_2_180a.eps,width=5.5cm}}}
\end{picture}
\hspace{-2.4cm}
\begin{picture}(200,200)  
\put(90,75){\makebox(10,20){\epsfig{figure=case1_2_180b.eps,width=5.5cm}}}
\end{picture}
\hspace{-2.4cm}
\begin{picture}(200,200)  
\put(110,75){\makebox(10,20){\epsfig{figure=case1_2_180c.eps,width=5.5cm}}}
\end{picture}
\caption{
Branching ratios for 
(a) $\tau^+ \to e^-e^+e^+$ and $\tau^+ \to \mu^-\mu^+\mu^+$,  
(b) $\tau^+\to e^-e^+\mu^+$  and $\tau^+\to \mu^-e^+e^+$, and  
(c) $\tau^+\to e^-\mu^+\mu^+$  and $\tau^+\to \mu^-e^+\mu^+$ 
against BR$(\mu^+\to e^+\gamma)$ for $\sin\theta_{13}=0.2$ with $\delta=\pi$ in CASE I.}
\label{fig:3}
\end{figure}

Using Eq.(\ref{h_coup}) the elements of $h_{ij}$ in CASE I are as follows: 
\beqa
&&-\sqrt{2}v_Rh_{e\mu}=c_{12}s_{12}c_{23}c_{13}(M_2-M_1)\pm c_{13}s_{13}s_{23}(M_3-s_{12}^{2}M_2-c_{12}^{2}M_1)\,, \label{h_emu}\\
&&-\sqrt{2}v_Rh_{e\tau}=
-c_{12}s_{12}s_{23}c_{13}(M_2-M_1)
\pm c_{23}c_{13}s_{13}(M_3-s_{12}^{2}M_2-c_{12}^{2}M_1)\,, \\
&&-\sqrt{2}v_Rh_{\mu\tau}=c_{23}s_{23}(c_{13}^{2}M_3-c_{12}^{2}M_2-s_{12}^{2}M_1)\pm c_{12}s_{12}s_{13}((s_{23}^{2}-c_{23}^{2})M_2+M_1) \nn \\
&&~~~~~~~~~~~~~~+c_{23}s_{23}s_{13}^{2}(s_{12}^{2}M_{2}-c_{12}^{2}M_1)\,,\\
&&-\sqrt{2}v_Rh_{ee}=c_{13}^{2}(c_{12}^{2}M_1+s_{12}^{2}M_2)+s_{13}^{2}M_3 \,,
\label{h_ee}\\
&&-\sqrt{2}v_Rh_{\mu\mu}=c_{23}^{2}(s_{12}^{2}M_1+c_{12}^{2}M_2)+s_{23}^{2}c_{13}^{2}M_3 
\mp 2c_{12}s_{12}c_{23}s_{23}s_{13}(M_2-M_1) \nn \\
&&~~~~~~~~~~~~~~+s_{23}^{2}s_{13}^{2}(c_{12}^{2}M_1+s_{12}^{2}M_2)\,.
\label{h_mumu}
\eeqa
Here the %+(-)
upper (lower) sign is for Dirac phase $\delta=0$ ($\pi$). 
It is clear that the off-diagonal elements in the 
$h$ couplings vanish when all 
the heavy neutrinos are degenerate in mass, i.e. $M_1=M_2=M_3$.
The strict bound on BR$(\mu\to eee)$ requires $|h_{e\mu}| \ll 1$ which
leads to the following conditions:
\beqa
&{\rm (i)}& M_1 \simeq M_2 , ~~~~~s_{13}(M_3-M_2) \simeq 0\ , 
%s_{13}(M_3-s_{12}^2M_2-c_{12}^2M_1) \simeq 0\ , 
\label{eq:CASEI_i}\\
&{\rm (ii)}& s_{13} M_3 \simeq c_{12}s_{12}(M_2-M_1)+s_{13}(s_{12}^2M_2+c_{12}^2M_1), ~~~~~\delta=\pi \ .
\label{eq:CASEI_ii}
%&(b)& s_{13}\simeq\frac{-c_{12}s_{12}c_{23}(M_2-M_1)}{s_{23}(-c_{12}^2M_1-s_{12}^2M_2+M_3)}\ne 0 \ , \\
%&   & (M_1\ne M_2 \to -c_1^2M_1-s_1^2M_2+M_3 >  0) \ .
\eeqa
In (i) both terms contributing to $h_{e\mu}$ are %exactly 
zero, 
while in (ii) there a cancellation which ensures $|h_{e\mu}| \ll 1$.

We show the branching ratios for all
six $\tau^+ \to lll$ decays against BR$(\mu^+ \to e^+ \gamma)$ for $s_{13}=0$ 
in Fig.\ref{fig:1}.
For simplicity we assume degeneracy
for the masses of the heavy  particles:
$M_{W_2}=M_{H_{L}^{\pm\pm}}=M_{H_{R}^{\pm\pm}}=M_{H^\pm_1}=3 {\rm ~TeV} $.
%Stabilization of the vacuum imposes stringent constraints 
%on the one-loop corrections
%to the effective potential and leads to the following relationship 
%\cite{Mohapatra:1986pj} between the masses for the $W_2$  boson 
%and the heavy neutrinos:
%\beq
%1.65M_{W_2} \geq M_{i} \ .
%\eeq
In our numerical analysis 
we vary the heavy neutrino masses randomly in the 
range 1 TeV $\leq M_i \leq$ 5 TeV, with distributions that are flat 
on a logarithmic scale. This range is consistent with
the vacuum stability condition for $v_R$ given in \cite{Mohapatra:1986pj}.

%Fig.\ref{fig:1} (a) shows the branching ratios for $\tau^+ \to e^+l_j^-l_k^-$ (l=e,$\mu$) 
%for which the couplings $h_{e\tau}$ and  $h_{l_jl_k}$ are relevant.
%The solid circles, plus signs, and open squares respectively denote the branching ratios for
%$\tau\to 3e$, $\tau^+\to e^-e^+\mu^+$, and $\tau^+\to e^-\mu^+\mu^+$.
%Fig.\ref{fig:1} (b) shows the branching ratios for $\tau^+ \to \mu^- l_jl_k$ (l=e,$\mu$)
%for which the couplings $h_{\mu\tau}$ and  $h_{l_jl_k}$ are relevant.
%The solid circles, plus signs, and open squares respectively denote the branching ratios for 
%$\tau^+\to \mu^- e^+e^+$, $\tau^+\to \mu^- e^+\mu^+$, and $\tau^+\to 3\mu$.
In Fig.\ref{fig:1} (a) ((b), (c)) the light and the dark points respectively denote the branching ratios for
$\tau^+ \to e^-e^+e^+$ and $\tau^+ \to \mu^-\mu^+\mu^+$ 
( $\tau^+\to e^-e^+\mu^+$ and $\tau^+\to \mu^-e^+e^+$,
$\tau^+\to \mu^-e^+\mu^+$ and $\tau^+\to e^-\mu^+\mu^+$).
We impose the experimental constraint BR$(\mu\to eee)<10^{-12}$ which prevents 
a large mass difference between $M_1$ and $M_2$ when $s_{13}=0$, as shown in
condition (i) of Eq.(\ref{eq:CASEI_i}). 
Among the six $\tau^+\to lll$ decay modes the branching ratios of 
$\tau^+\to \mu^-\mu^+\mu^+$ and $\tau^+\to \mu^- e^+e^+$
can reach the anticipated sensitivity 
($> 5\times 10^{-9}$) of a future B factory. 
Such branching ratios are realized when %$M_3$ 
the mass splitting $M_3-M_2$ assumes larger values because 
$h_{\mu\tau}$ increases with it. %the mass splitting $M_3-M_2$.
On the other hand, the $\tau\to e$ transition is suppressed because 
$|h_{e\mu}|=|h_{e\tau}|$ for $s_{13}=0$, and $|h_{e\mu}|$ 
is necessarily small in order to comply with the
severe constraint from $\mu\to eee$.
This can be seen for the light points in Fig.\ref{fig:1} (a) 
where BR$(\tau \to eee)$ is proportional to BR$(\mu\to eee)$.
Moreover, there is a strong correlation  
BR$(\tau\to eee)\sim 10\times {\rm BR}(\mu\to e\gamma)$.
BR($\mu\to e\gamma$) can be large as $10^{-14}$, which is 
within the sensitivity of MEG experiment.

Figs.\ref{fig:2} and \ref{fig:3} show the branching ratios for 
$\tau \to lll$ for $s_{13}=0.2$ with the Dirac phase $\delta=0$ 
and $\pi$ respectively, imposing the constraint BR$(\mu\to eee)<10^{-12}$.
For the other parameters we take the same values as in Fig.\ref{fig:1}.
When $\delta=0$ (Fig.\ref{fig:2}) all LFV processes are predicted 
to be small because of the small mass differences between the 
heavy neutrinos as shown in the condition (i).
However, there is still the possibility of 
observing $\mu\to e \gamma$ at MEG experiment. 
On the other hand, taking $\delta=\pi$ 
(Fig.\ref{fig:3}) results in observable 
branching ratios for $\tau\to lll$.
In this scenario one has $|h_{e\mu}|<|h_{e\tau}|$.
Therefore the equality of $|h_{e\mu}|$ and $|h_{e\tau}|$ in Fig.\ref{fig:1}
is broken, which enables enhancement of 
BR($\tau^+ \to e^-e^+e^+$) and BR($\tau^+ \to e^-\mu^+\mu^+$) 
with simultaneous suppression of the $\mu \to e$ transition.
Moreover, BR$(\mu\to e\gamma)$ can be large, resulting in 
multiple signals of LFV processes.

%%%%%%%%%%%%%%%%%%%%%%%%%%%%%%%%%%%%
\subsection{Numerical results: CASE II }
\begin{figure}[ht]
\centering
\begin{picture}(200,200)  
\put(45,75){\makebox(100,20){\epsfig{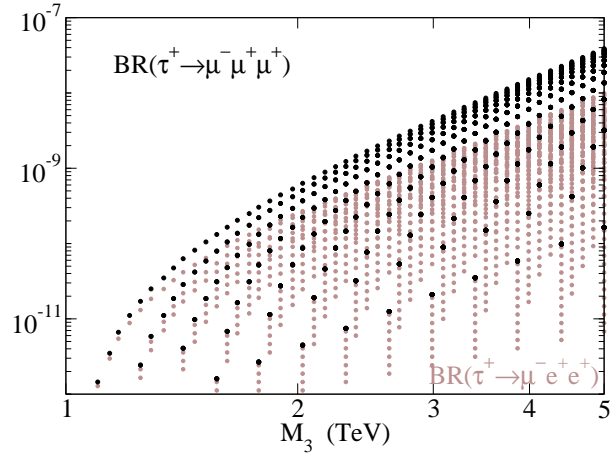}}}
\end{picture}~~
\caption{
Branching ratios for $\tau^+ \to \mu^-\mu^+\mu^+$ and $\tau^+ \to \mu^- e^+e^+$ 
against $M_3$ for $\sin\theta_{13}=0$ in CASE II.}
\label{fig:4}
\end{figure}

%%%%%%%%%%%%%%%%%%%%%%%%%%%%%%%%%%%%
In this case the large mixing for the atmospheric angle originates
from the charged lepton sector, while the large solar angle originates
from the neutrino sector. The $R$ matrix (given in Eq.(\ref{eq:Rmatrix})) 
which satisfies the 
condition $m_D=m_D^\dag$ is:
\beqa
\tan\theta_{12}^R&=&\frac{\sqrt{M_1m_1}+\sqrt{M_2m_2}}{\sqrt{M_1m_2}+\sqrt{M_2m_1}}\tan\theta_{12} \ ,\\
\theta_{23}^R&=&\theta_{13}^R=0\ .
%\sin\theta_{23}^R&=&0,~~\sin\theta_{13}^R=0\ .
\eeqa
The explicit form for $h_{ij}$ is obtained from 
Eq.(\ref{h_emu}) - (\ref{h_mumu}) by taking $s_{12}=0$.
The condition for suppressing $|h_{e\mu}|$ is:
\beq
s_{13}(M_3-c_{12}^2M_1) \simeq 0 \ ,
\label{II_i}
\eeq
which requires $s_{13}\simeq 0$ or $M_1\simeq M_2\simeq M_3$.
In the latter case none of the $\tau\to lll$ decays are measurable,
as in CASE I with $s_{13}=0.2$ and $\delta=0$. On the other hand, 
$h_{e\mu}$ and $h_{e\tau}$ are zero when $s_{13}=0$, 
which results in vanishing branching ratios for $\tau^+ \to e^-l^+l^+, \mu^- e^+ \mu^+$
and $\mu\to e\gamma$.
Fig.\ref{fig:4} shows the branching ratios for 
$\tau^+ \to \mu^-\mu^+\mu^+$ and $\tau^+ \to \mu^- e^+e^+$ 
against the heaviest neutrino mass 
$M_3$. Clearly the branching ratios increase with $M_3$ and
for $M_3>3$ TeV observable rates are attained.

%%%%-%%%%%%%%%%%%%%%%%%%%%%%%%%%%%%%%
%    Figure 4" (CASE II,  s3 = 0)
%%%%%%%%%%%%%%%%%%%%%%%%%%%%%%%%%%%%%%%

%%%%%%%%%%%%%%%%%%%%%%%%%%%%%%%%%%%%
%
%    Figure 5  (CASE II,  s3 \ne 0)
%%%%%%%%%%%%%%%%%%%%%%%%%%%%%%%%%%%%%%%
\begin{figure}[ht]
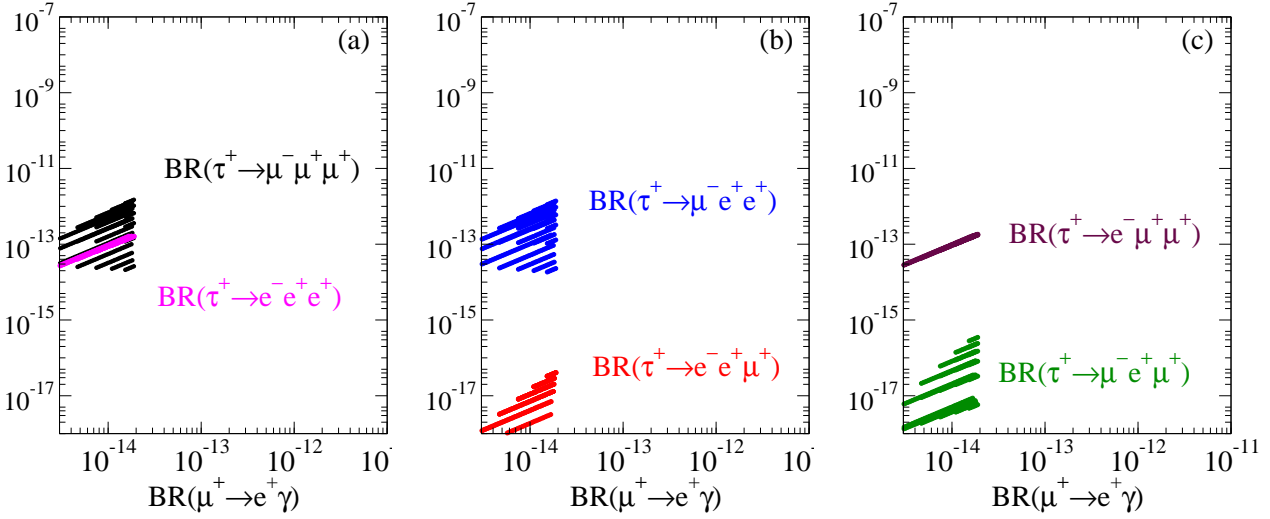

\centering
\begin{picture}(200,200)  
\put(70,75){\makebox(10,20){\epsfig{figure=case2_2a.eps,width=5.5cm}}}
\end{picture}
\hspace{-2.4cm}
\begin{picture}(200,200)  
\put(90,75){\makebox(10,20){\epsfig{figure=case2_2b.eps,width=5.5cm}}}
\end{picture}
\hspace{-2.4cm}
\begin{picture}(200,200)  
\put(110,75){\makebox(10,20){\epsfig{figure=case2_2c.eps,width=5.5cm}}}
\end{picture}
\caption{
Branching ratios for 
(a) $\tau^+ \to e^-e^+e^+$ and $\tau^+ \to \mu^-\mu^+\mu^+$,  
(b) $\tau^+\to e^-e^+\mu^+$  and $\tau^+\to \mu^-e^+e^+$, and  
(c) $\tau^+\to e^-\mu^+\mu^+$  and $\tau^+\to \mu^-e^+\mu^+$ 
against BR$(\mu\to e\gamma)$ for $s_{13}=0.2$ with $\delta=0,\pi$ in CASE II.}
\label{fig:5}
\end{figure}

When $s_{13}\ne 0$ the magnitudes of each entry in 
$h_{ij}$ are the same for $\delta=0$ and $\pi$.
We show in Fig.\ref{fig:5}
the branching ratios for $\tau\to lll$ decay 
against BR$(\mu\to e\gamma)$ 
for $s_{13}=0.2$.
In order to satisfy the condition in 
Eq.(\ref{II_i}) there cannot be large splittings
among the masses of the heavy neutrinos. 
In this scenario only BR$(\mu\to e\gamma)$ reaches
the future experimental sensitivity.   

%%%%%%%%%%%%%%%%%%%%%%%%%%%%%%%%%%%%
\subsection{Numerical Results: CASE III}
%%%%%%%%%%%%%%%%%%%%%%%%%%%%%%%%%%%%
In CASE III the constraint from 
$\mu\to eee$ is satisfied automatically
since $h_{e\mu}=0$ (obtained by setting $s_{12}=s_{13}=0$ in 
Eq.(\ref{h_emu}).
We obtain $R$ as follows by treating $\theta_{13}$ as a perturbation:
% ($\cos\theta_{13}^R\sim 1-(1/2)\sin^2\theta_{13}^R$, ).
\beqa
\tan\theta_{12}^R\hspace{-3mm}&=&\hspace{-3mm}
\frac{\sqrt{M_1m_1}+\sqrt{M_2m_2}}{\sqrt{M_1m_2}+\sqrt{M_3m_1}}\tan\theta_{12}
\ , \\
\sin\theta_{23}^R\hspace{-3mm}&=&\hspace{-3mm}\frac{g}{f}s_{13} \ , \\
f\hspace{-3mm}&=&\hspace{-3mm}M_3\sqrt{m_1m_2}+\sqrt{M_3m_3}\left\{ \left(\sqrt{M_1m_1}+\sqrt{M_2m_2}\right)s_{12}s_{12}^R+\left(\sqrt{M_1m_2}+\sqrt{M_2m_1}\right)c_{12}c_{12}^R  \right\} \nn \\
&&+\sqrt{M_1M_2}m_3 \ , \\
g\hspace{-3mm}&=&\hspace{-3mm}\sqrt{M_3}c_{12}s_{12}\left\{\left(\sqrt{M_2m_1m_2}+\sqrt{M_1}m_1\right)c_{12}^{R2}-\left(\sqrt{M_2m_1m_2}+\sqrt{M_1}m_2\right)s_{12}^{R2}  \right\} \nn \\
&&+\left\{\sqrt{M_2M_3}(m_2s_{12}^2-m_1c_{12}^2)+\sqrt{M_1M_3m_1m_2}(s_{12}^2-c_{12}^2) \right\}
c_{12}^Rs_{12}^R  \nn \\
&&+\sqrt{m_3}\left\{\left(\sqrt{M_1M_2m_2}+M_3\sqrt{m_1}\right)s_{12}c_{12}^R-\left(\sqrt{M_1M_2m_1}+M_3\sqrt{m_2}\right)c_{12}s_{12}^R  \right\} \ , \\
\sin\theta_{13}^R\hspace{-3mm}&=& \nn \\ %\hspace{-3mm}
\hspace{-1.2cm}&&\hspace{-1.2cm}
\frac{\sqrt{M_3}(\sqrt{m_1}c_{12}s_{12}^R-\sqrt{m_2}s_{12}c_{12}^R)s_{23}^R+
\left\{\sqrt{M_3m_3}+\sqrt{M_1}(\sqrt{m_1}c_{12}c_{12}^R+\sqrt{m_2}s_{12}s_{12}^R) \right\}s_{13}   }
{\sqrt{M_1m_3}+\sqrt{M_3}(\sqrt{m_1}c_{12}c_{12}^R+\sqrt{m_2}s_{12}s_{12}^R)}\ .
\eeqa

CASE III predicts the same results as CASE II for the branching ratios of 
the LFV processes with $s_{13}=0$ (Fig.\ref{fig:4}).

%%%%%%%%%%%%%%%%%%%%%%%%%%%%%%%%%%%%

\subsection{Numerical results: CASE IV}
%%%%%%%%%%%%%%%%%%%%%%%%%%%%%%%%%%%%
In this case the bi-large mixing originates from the neutrino sector. 
We checked the existence of $R$ numerically.
It is clear that none of the LFV processes are observed since 
$h_{ij}$ is a diagonal matrix.

\subsection{Summary}

The numerical results of CASES I, II, III and IV 
for $M_{W_2}=M_{H^{\pm\pm}_L}=M_{H^{\pm\pm}_R}=M_{H_1^\pm}=3$ TeV
are qualitatively summarized in Table 1.
\begin{table}
{\footnotesize
\begin{center}
\caption{LFV processes within the sensitivity of forthcoming or 
planned experiments for CASES I, II, III and IV 
with $M_{W_2}=M_{H^{\pm\pm}_L}=M_{H^{\pm\pm}_R}=M_{H_1^\pm}=3$ TeV.
BR($\mu\to e\gamma)> 10^{-14}$ and BR($\tau\to lll)> 10^{-9}$ 
is denoted by ``$\surd$''.}
\begin{tabular}{@{\vrule width0.8pt}c|ll
@{\quad\vrule width0.8pt} c|cc|cc@{\quad\vrule width0.8pt}}
\noalign{\hrule height 1pt}
CASE  &$\sin\theta_{13}$ &$\delta$ &$\mu^+ \to e^+\gamma$ &$\tau^+ \to e^-e^+e^+ $&$\tau^+ \to e^-\mu^+\mu^+ $
&$\tau^+\to \mu^- e^+e^+$ &$\tau^+\to \mu^-\mu^+\mu^+$ \\
%&$\tau^+\to e^+ e^-\mu^-$ &$\tau^+\to \mu^+ e^-\mu^-$  \\
\noalign{\hrule height 1pt}
& $0$         &           &  $\surd$  &           &             & $\surd$ &$\surd$  \\
I& 0.2,&0   & $\surd$  &             &             &            &              \\
& 0.2,&$\pi$  &  $\surd$ & $\surd$ & $\surd$ & $\surd$ &$\surd$    \\
\noalign{\hrule height 1pt}
& 0     &              &            &           &             & $\surd$ &$\surd$    \\
II& 0.2,&0   &   $\surd$ &             &             &            &               \\
& 0.2,&$\pi$  &   $\surd$ &            &             &            &               \\
\noalign{\hrule height 1pt}
& 0      &           &            &           &             & $\surd$ &$\surd$    \\
III& 0.2,&0   &            &           &             & $\surd$ &$\surd$   \\
& 0.2, &$\pi$  &           &           &             & $\surd$ &$\surd$    \\
\noalign{\hrule height 1pt}
& 0       &         & &&&& \\
IV& 0.2,&0   &&&&& \\
& 0.2,&$\pi$  &&&&& \\
\noalign{\hrule height 1pt}
\end{tabular}
\end{center}
}
\label{result_sum}
\end{table}
It is clear that the number of observable rates 
for the LFV decays $\tau\to lll$ and $\mu\to e\gamma$ depends
sensitively on the origin of 
the bi-large mixing in $V_{\rm MNS}$, with up to five signals being
possible in the optimum scenario (CASE I with $\sin\theta_{13}=0.2$,
$\delta=\pi$).
Hence future searches for $\tau\to lll$
and $\mu\to e\gamma$ provide insight into the structure of
$h_{ij}$ in the LR model.

Both BR($\tau\to lll$) and BR($\mu\to eee$) are inversely
proportional to the fourth power of 
$M_{H^{\pm\pm}}$ (when $M_{H^{\pm\pm}} 
=M_{H^{\pm\pm}_L}=M_{H^{\pm\pm}_R}$) as shown in Eq.(\ref{BRtaulll}).
Reducing $M_{W_2}$, $M_{H^{\pm\pm}_L}$, and $M_{H^{\pm\pm}_R}$
in all the figures would cause enhancement of BR($\tau\to lll$)
and BR($\mu\to eee$) while maintaining the correlation.
However, BR($\mu\to eee$) vanishes in CASE II with $s_{13}=0$ and 
CASE III, and hence the strong bound BR($\mu\to eee)<10^{-12}$ 
is automatically satisfied for any $M_{H^{\pm\pm}}$.
In these latter cases the current experimental limits
for BR($\tau\to lll$) can be reached and
thus an upper bound on the heaviest neutrino mass $M_3$
can be derived 
({\it e.g.} Fig.\ref{fig:4} with $M_{W_2}=M_{H^{\pm\pm}}=1.5$ TeV
gives BR($\tau\to \mu\mu\mu) > 10^{-7}$ for $M_3 > 1.8$ TeV). 

We note here that the presented numerical results are
for the scenario of all phases in $V_{\rm MNS}$ taken to be zero.
We now briefly discuss the effect of including a non-zero Dirac 
phase ($\delta\ne 0$) in our analysis
\footnote{ The presence of Majorana phases in $V_{\rm MNS}$ would effectively
change the relative phase of the heavy neutrino masses $M_i$ 
in Eqs.(41) to (45). In such a scenario small $|h_{ee}|$ can arise
from a cancellation among the terms in Eq.(44), which 
provides an additional way to suppress BR$(\mu \to eee)$ in CASE I.
We performed an explicit numerical analysis 
and found a pattern of lepton flavour violation different
from the case of $|h_{e\mu}|\sim 0$. In particular, 
BR($\tau^+ \to e^-e^+\mu^+$) and BR($\tau^+ \to \mu^-e^+\mu^+$) 
can be enhanced to observable rates. We thank the referee for
bringing this scenario to our attention.}.
In CASE I and CASE II the presence of $\delta\ne 0$ 
would not affect 
Eq.(\ref{eq:mD}) since $U(\theta_{13}$) in 
Eq.(\ref{eq:MNS}) (which contains $\delta$)
only appears in $V_L^l$ and not in $V_L^\nu$.
Thus the solution for the $R$ matrix in the CP conserving case
($\delta=0$ or $\pi$) also holds for the case of $\delta\ne 0$.
Since $V^l_L$ (which determines the LFV rates) 
has some dependence on $\delta$ we would expect some
%minor 
changes in our numerical results.
For example, we have calculated the branching ratios for the LFV processes for $\delta=\pi/2$ and $3\pi/2$ 
with $s_{13}=0.2$ in CASE I and II. For all cases BR$(\mu\to e\gamma) < 2\times 10^{-14}$.
 Maximum values of $10^{-9}$ were found for
 BR$(\tau^+\to\mu^-\mu^+\mu^+ )$,  BR$(\tau^+\to\mu^-
e^+e^+ )$, and  BR$(\tau^+\to e^-\mu^+\mu^+ )$,
with smaller ($<10^{-10}$) values for
BR$(\tau^+\to e^-e^+e^+ )$,  BR$(\tau^+\to e^- e^+\mu^+ )$ and
BR$(\tau^+\to \mu^- e^+\mu^+ )$.
When $\delta \ne 0$ there is another possibility to suppress BR$(\mu\to eee)$
by  cancellation in $h_{ee}$, because the sign of the term $s_{13}^2M_3$
in Eq.(\ref{h_ee}) flips for $\delta=\pi/2$ and $3\pi/2$. However, the
cancellation is not so significant 
because of the small factor $s_{13}^2$.
% but the same qualitative conclusions. Although $\delta\ne 0$ can change the sign of 
%the contribution from $s^2_{13}M_3$ in Eq.(\ref{h_ee}), the small factor
%$s^2_{13}$ prevents sufficient cancellation to enable $h_{ee}\simeq 0$.
%Hence even for $\delta\ne 0$ the necessary suppression of BR($\mu\to eee$) 
%requires $h_{e\mu}\simeq 0$.
In CASE III and IV the presence of $\delta\ne 0$ increases the
number of free parameters in $V_L^\nu$ and no
general solution for the $R$ matrix can be found.
However we expect that Eq.(\ref{eq:mD_hermitian}) will
be satisfied in specific regions in the parameter space of $M_i$,
and for CASE III some observable LFV rates might still be possible.

\section{$P$ odd Asymmetry for $\tau\to l_il_jl_k$ and $\mu\to e\gamma$}
%\section{Distinguishing models
% with $H^{\pm\pm}$ via
%$P$ odd Asymmetry for $\tau\to l_il_jl_k$ and $\mu\to e\gamma$}

From the preceding sections it is evident that both 
BR$(\tau\to lll$) and BR($\mu\to e\gamma$)
can be enhanced to experimental observability in the LR symmetric model. 
In particular, BR$(\tau\to lll)>10^{-8}$ would be a signal suggestive of
models which can mediate the decay at tree-level  
e.g. the LR model via virtual exchange of $H^{\pm\pm}_{L,R}$
\footnote{For models with large extra dimensions see \cite{Chang:2005ag}.}.
In order to compare the LR model with other models we introduce 
the Higgs Triplet Model and Zee-Babu Model in section 4.1 and 4.2.
These models provide a different mechanism for neutrino mass generation,  
and can enhance both $\tau\to l_il_jl_k$ and $\mu\to e\gamma$ to
experimental observability by virtual exchange of $H^{\pm\pm}_L$
or $H^{\pm\pm}_R$.
If a signal were established for any of the six decays $\tau\to lll$,
further information on the underlying model 
can be obtained by studying the angular distribution of the leptons.
%In this section we show that asymmetries can be defined which
%may disciminate models which possess $H^{\pm\pm}$. 
In section 4.3 we show how the
$P$ odd asymmetry for both decays may act as a powerful discriminator of
the three models under consideration.

\subsection{Higgs Triplet Model}

In the Higgs Triplet Model (HTM) \cite{Schechter:1980gr}
a single $I=1$, $Y=2$ complex $SU(2)_L$ triplet is added to the SM.
No right-handed neutrino is introduced, and 
the light neutrinos receive a Majorana mass proportional
to the left-handed triplet vev $(v_L)$
leading to the following neutrino mass matrix:
\begin{equation}
{\cal M}_{\nu}=\sqrt{2}v_{L}
\left( \begin{array}{ccc}
h_{ee}  & h_{e\mu} & h_{e\tau} \\
h_{\mu e}  & h_{\mu\mu} & h_{\mu\tau} \\
h_{\tau e}  & h_{\tau \mu} & h_{\tau\tau}
\end{array} \right)  \; .
\end{equation}
In the HTM $h_{ij}$ is directly related to the neutrino
masses and mixing angles as follows:
\begin{equation}
h_{ij}=\frac{1}{\sqrt{2}v_L}V_{_{\rm MNS}}diag(m_1,m_2,m_3)
V_{_{\rm MNS}}^T\ .
\label{hij}
\end{equation}
Formally, this expression for $h_{ij}$ is equivalent to that in 
CASE I with the replacements $(m_1,m_2,m_3)\to (M_1,M_2,M_3)$
and $v_L\to v_R$. In Eq.(\ref{hij}) $v_L$
is a free parameter and is necessarily non-zero 
(unlike $v_L$ in the LR symmetric model) in order to
generate neutrino masses. Its magnitude may lie anywhere 
in the range eV $< v_{L} <$ 
8 GeV, where the lower limit arises from the requirement
of a perturbative $h_{ij}$ satisfying Eq.(\ref{hij})
and the upper limit is derived from 
maintaining $\rho\sim 1$. In the HTM 
there is no $H^{\pm\pm}_R$ and so $\tau\to l_il_jl_k$
is mediated solely by $H^{\pm\pm}_L$.
Obtaining BR($\tau\to l_il_jl_k)>10^{-8}$ with $m_{H^{\pm\pm}}= 1 $ TeV
requires $|h_{\tau i}^\ast h_{jk}|>10^{-3}$.
Due to the ignorance of $v_L$ the magnitude of $h_{ij}$ cannot be predicted
and so the HTM can only {\it accommodate} observable BRs 
for $\tau\to l_il_jl_k$.
However, unlike the LR symmetric model, the HTM provides predictions 
for the ratios of $\tau\to l_il_jl_k$ \cite{Chun:2003ej} with are distinct
for each of the various neutrino mass patterns NH, IH and DG. 
The necessary suppression of BR$(\mu\to eee$) relative to
BR$(\tau\to lll$) requires  
$h_{e\mu}$ to be sufficiently small. As in CASE I Eq.({\ref{eq:CASEI_ii}}),
this is arranged by invoking
a cancellation between two terms contributing to $h_{e\mu}$, 
one depending on $\theta_{13}$ and the other
depending on both $\theta_{12}$ and 
$r=\Delta m^2_{12}/\Delta m^2_{13}$ \cite{Chun:2003ej},\cite{Kakizaki:2003jk}.
Observation of $\tau\to lll$ would
restrict $\theta_{13}$ to a narrow interval which can be predicted
in terms of $\theta_{12}$ and $r$. 
For the case of NH neutrinos $\theta_{13}\sim \sqrt r$
while for IH and DG neutrinos $\theta_{13}\sim r$.
Since current oscillation data suggests $r \approx 0.04$, 
the value of $\theta_{13}$ need to ensure small $h_{e\mu}$ 
is of the order $0.1$ for NH and $0.01$ for IH and DG.

\subsection{Two loop radiative singlet Higgs model (Zee-Babu model)} 
Neutrino mass may be absent at the tree-level but is generated radiatively
via higher order diagrams involving $L=2$ scalars.
In the Zee-Babu model (ZBM)
$SU(2)_L$ singlet charged scalars $H^{\pm\pm}_R$ and $H^\pm_L$ are added
to the SM Lagrangian  \cite{Zee:1985id},\cite{Babu:2002uu}
with the following Yukawa couplings:
\begin{equation}
{\cal L}_{Y} = f_{ij}\left( L^{T a}_{iL} C L^b_{jL} \right)
\epsilon_{ab} H^+_L \
+ \ h'_{ij} \left( l^T_{iR} C l_{jR} \right) H^{++}_R + h.c. \ 
\end{equation}
No right-handed neutrino is
introduced. A Majorana mass for the light neutrinos 
arises at the two loop level in which the 
lepton number violating trilinear coupling 
$\mu H^\mp_L H^\mp_L H^{\pm\pm}_R$ plays a crucial role. 
The explicit form for ${\cal M}_{\nu}$ is as follows:
{\normalsize
\beq
  {\cal M}_{\nu} = \zeta
  \times \left(
\begin{array}{ccc}
\epsilon^2 \omega_{\tau \tau} + 2 \epsilon \epsilon' \omega_{\mu \tau} + 
\epsilon'^2 \omega_{\mu \mu}\ , &
\epsilon \omega_{\tau \tau}  + \epsilon' \omega_{\mu \tau} - 
\epsilon \epsilon' \omega_{e \tau}
  &
-\epsilon \omega_{\tau\tau} -\epsilon' \omega_{\mu\mu} - \epsilon^2 
\omega_{e\tau}  \\
 &  \hbox{~~~~~~~~~} - \epsilon'^2 \omega_{e \mu} \ , &
 \hbox{~~~~~~~~~} - \epsilon \epsilon' \omega_{e\mu} \\
%-----end of line 1
. & \omega_{\tau\tau} -2 \epsilon' \omega_{e\tau} + \epsilon'^2 
\omega_{ee} \ , &
-\omega_{\mu\tau} -\epsilon \omega_{e\tau} + \epsilon' \omega_{e\mu}  \\
 & & \hbox{~~~~~~~~~} + \epsilon \epsilon' \omega_{ee} \\
%-----end of line 2
. & . & \omega_{\mu\mu} + 2 \epsilon \omega_{e\mu} 
+ \epsilon^2 \omega_{ee} 
\end{array}
\right)\ ,
\eeq
}where $\epsilon=f_{e\tau}/f_{\mu\tau}$, $\epsilon'=f_{e\mu}/f_{\mu\tau}$,
$\omega_{ij}=h_{ij}m_i m_j$ ($m_i,m_j$ are charged fermion masses),
$h_{ij}=h'_{ij}(2h'_{ij})$ for $i=j$ ($i\ne j$)  
and $\zeta$ is given by:
\begin{equation}
\zeta=\frac{8\mu f^2_{\mu\tau}\tilde I}{(16\pi^2)^2m_{H^\pm}^2} \ .
\end{equation}
Here $\tilde I$ is a dimensionless quantity of ${\cal O}(1)$
originating from the loop integration.
Clearly the expression for ${\cal M}_{\nu}$ differs from that in the 
HTM and involves 9 arbitrary couplings.
Since the model predicts one massless neutrino (at the two-loop level),
quasi-degenerate neutrinos are not permitted (unlike the HTM)
and only NH and IH mass patterns can be accommodated.
The $f$ couplings (contained in 
$\epsilon$ and $\epsilon'$) are directly related to the elements
of ${\cal M}_{\nu}$.
In the scenario of NH, $\epsilon\approx \epsilon'\approx 
\tan\theta_{12}/\sqrt 2$
and $\sin\theta_{13}$ is close to zero. Since $\epsilon,\epsilon'<1$
one may neglect those terms in ${\cal M}_{\nu}$ which 
are proportional to the electron mass (i.e. $\omega_{ee},\omega_{e\mu},
\omega_{e\tau}$).
This simplification leads to the following prediction
\cite{Babu:2002uu}: $h_{\mu\mu}:h_{\mu\tau}:
h_{\tau\tau}\approx 1:m_\mu/m_\tau:(m_\mu/m_\tau)^2$.
In the case of IH, large values  
are required for $\epsilon,\epsilon'(>5)$, and thus neglecting
  $\omega_{ee},\omega_{e\mu},
\omega_{e\tau}$ in ${\cal M}_{\nu}$
may not be entirely justified. However, if such terms are neglected
\cite{Babu:2002uu} then the above prediction for the ratio
of $h_{\mu\mu}:h_{\mu\tau}:h_{\tau\tau}$ also approximately holds
for the case of IH. In the ZBM there is no $H^{\pm\pm}_L$ and so
$\tau\to l_il_jl_k$ is mediated solely by $H^{\pm\pm}_R$. 

Another significant difference with the HTM is that
eV scale neutrino masses requires $f$, $h_{\mu\mu}\sim 10^{-2}$,
and thus LFV decays cannot be suppressed arbitrarily if
the 2-loop diagram is solely responsible for the generation
of the neutrino mass matrix.
Such relatively large couplings are necessary since 
a rough upper bound on $\zeta$ (which is a function of model parameters)
can be derived. In contrast, $v_L$ in the HTM is arbitrary and
eV scale neutrino masses can be accommodated even with
$h_{ij}\sim 10^{-10}$.
The requirement that $f$, $h_{\mu\mu}\sim 10^{-2}$ suggests 
that BR($\mu\to e\gamma$) and BR$(\tau\to \mu\mu\mu$) could be
within range of upcoming experiments \cite{AristizabalSierra:2006gb}.

Since $h_{ee}$, $h_{e\mu}$ and $h_{e\tau}$ may be
treated as free parameters (essentially unrelated to the neutrino
mass matrix) the necessary suppression of $\mu\to eee$ can be obtained 
by merely choosing $h_{e\mu}$ and/or $h_{ee}$ very small. 
Observable rates for BR($\tau\to eee$) can be arranged by appropriate
choice of $h_{ee}$ and  $h_{e\tau}$.

\subsection{Sensitivity of $P$ odd asymmetry to $H^{\pm\pm}_L$ and 
$H^{\pm\pm}_R$}
Angular distributions of LFV decays can act as a powerful discriminator
of models of new physics. 
The predictions for 
$\mu^+\to e_L^+\gamma$ and $\mu^+\to e_R^+\gamma$ depend on 
the chirality structure of LFV interactions and so in general
would be model dependent. 
Ref.\cite{Okada:1999zk} defined various $P$ odd and $T$ odd asymmetries
for $\mu\to e\gamma$ and $\mu\to eee$ and performed
a numerical analysis in the context of supersymmetric $SU(5)$ and $SO(10)$. 
Analogous asymmetries were defined for $\tau^\pm\to l^\pm\gamma$ and
$\tau^\pm\to lll$ in \cite{Kitano:2000fg}.
In this section we apply the general formulae introduced
in Refs.\cite{Kitano:2000fg},\cite{Okada:1999zk}
 to the three models of interest which all contain $H^{\pm\pm}$.

For the decay $\mu^+\to e^-e^+e^+$ with polarized $\mu^+$, 
one defines $\theta_{e^-}$ as
the angle  between the polarization vector of $\mu^+$
and the direction of the $e^-$, the latter taken to be
the $z$ direction. 
%The helicity of $\mu^+$ is determined by the experimental set up,
%and hence the angle $\theta$ can be immediately defined
%once a signal for $\mu^+\to e^+e^+e^-$ is observed. 
The $P$ odd
asymmetry ${\cal A}_{P}$ is defined as 
an asymmetry in the $\theta_{e^-}$ distribution.
In contrast, for $\tau$ produced in the process $e^+e^-\to \tau^+\tau^-$
the helicity of the $\tau$ in the LFV decay $\tau\to lll$ 
is not known initially. Consequently, the experimental set up 
is sensitive to both $\tau^+_L\to lll$ or $\tau^+_R\to lll$.
However, by exploiting the spin correlation of the 
pair produced $\tau$  (i.e. $e^+e^-\to \tau^+_L\tau^-_R,\tau^+_R\tau^-_L$)
information on the helicity of the LFV decaying $\tau$
can be obtained by studying the angular and kinematical
distributions of the non-LFV decay 
of the other $\tau$ in the $\tau\to lll$ event. For illustration we
shall always take the non-LFV decay mode as $\tau\to \pi\nu$, although
such an analysis can also be performed for other main decay modes such as
$\tau\to \rho\nu,a_1\nu,l\nu\overline \nu$.

In the notation of \cite{Kitano:2000fg} the effective
4-Fermi interaction for %$\tau^+\to l_i l_j l_k$ 
$\tau^+\to \mu^-\mu^+\mu^+$ mediated
by $H^{++}_{L,R}$ is as follows:
\begin{equation}
{\cal L}=\frac{-4G_F}{\sqrt 2}\left\{g_3
(\overline \tau\gamma^\mu P_R\mu)(\overline\mu\gamma_\mu P_R\mu)
+g_4(\overline \tau\gamma^\mu P_L\mu)(\overline\mu\gamma_\mu P_L\mu)\right\}\ . %+ h.c
\end{equation}
%Two $P$-odd asymmetries (${\cal A}_{p_1}$ and ${\cal A}_{p_2}$)
%can be defined for $\tau\to lll$
%which differ in the integration region for $\theta$.
%In the LR model only $g_3$ and $g_4$ appear in the effective Lagrangian
%which leads to ${\cal A}_{p_2}=0$, while in general ${\cal A}_{p_1}\ne 0$.
The differential cross-section for the events $\tau^+\to \mu^-\mu^+\mu^+$ (LFV)
and $\tau^-\to \pi^-\nu$ (non-LFV) is:
\begin{eqnarray}
 &&d\sigma (e^+ e^- \to \tau^+ \tau^- \to \mu^- \mu^+ \mu^+ + \pi^- \nu)
\nonumber \\
&& = \sigma (e^+ e^- \to \tau^+ \tau^-) 
BR (\tau^- \to \pi^- \nu) \left(
\frac{m_\tau^5 G_{\rm F}^2}{128 \pi^4}/\Gamma
\right)
\frac{d \cos \theta_\pi}{2}\ dx_1 \ dx_2\ d\cos \theta \ d\phi
\nonumber \\
&& \hspace*{1cm} \times
\left[
X - \frac{s-2m_\tau^2}{s+2 m_\tau^2} 
\left \{ 
Y \cos \theta
\right \}
\cos \theta_\pi 
\right] \ ,
\label{eq.taulll_cross}
\end{eqnarray}
where 
\begin{equation}
X=(|g_3| ^2+|g_4|^2)\alpha_1(x_1,x_2);\,\,Y=(|g_3| ^2-|g_4|^2)\alpha_1(x_1,x_2) \ .
\end{equation}
Here $\alpha_1(x_1,x_2)$ is a function of the 
energy variables $x_1=2E_1/m_{\tau}$ and
$x_2=2E_2/m_{\tau}$ where $E_1(E_2)$ is
the energy of $\mu^+$ with larger (smaller) energy 
in the rest frame of $\tau^+$:
\beq
 \alpha_1 (x_1, x_2) = 8(2-x_1-x_2)(x_1+x_2-1) \ .
\eeq
The angles $\theta$ and $\phi$ specify the decay plane of 
$\tau^+\to \mu^-\mu^+\mu^+$ relative to the production plane of
$e^+e^-\to \tau^+\tau^-$ in the $\tau^+$ rest frame.
The angle $\theta_\pi$ is the angle between 
the direction of momentum of $\tau^-$ and $\pi^-$
in the $\tau^-$ rest frame. For a detailed discussion 
we refer the reader to \cite{Kitano:2000fg}.
It is clear that $Y$ determines the angular dependence of
Eq.(\ref{eq.taulll_cross}) and thus $Y$ is a measure of 
the $P$ odd asymmetry for $\tau\to \mu\mu\mu$.
For our quantitative study of its magnitude we define:
\begin{equation}
{\cal A}(\tau\to \mu\mu\mu)=\frac{|g_3|^2-|g_4|^2}{|g_3|^2+|g_4|^2} \ .
\end{equation}
Clearly ${\cal A}(\tau\to \mu\mu\mu)=0$ $(\pm 1)$ corresponds to zero (maximal)
asymmetry.
\begin{table}
\begin{center}
\caption{Expressions for $g_3$, $g_4$ in the three models}
%\begin{minipage}[t]{0.8\textwidth}
\begin{tabular}{cccc}
\hline
 & HTM ($H^{\pm\pm}_L$) & ZBM ($H^{\pm\pm}_R$) &  LR ($H^{\pm\pm}_{L,R}$) \\
\hline
&&&\\[-3.5mm]
$-\frac{4G_F}{\sqrt{2}}g_3$ & 0 & $\frac{h_{\mu\mu}h_{\tau \mu}^\ast}{M^2_{H^{\pm\pm}_R}}$
 & $\frac{h_{\mu\mu}h_{\tau \mu}^\ast}{M^2_{H^{\pm\pm}_R}}$  \\[3.5mm]
$-\frac{4G_F}{\sqrt{2}}g_4$  &  $\frac{h_{\mu\mu}h_{\tau \mu}^\ast}{M^2_{H^{\pm\pm}_L}}$  
& 0 &   $\frac{h_{\mu\mu}h_{\tau \mu}^\ast}{M^2_{H^{\pm\pm}_L}}$  
  \\[4mm]  \hline
\end{tabular}
%\end{minipage}
\end{center}
\end{table}

\begin{figure}[h]
\begin{center}
\includegraphics[width=6.5cm,angle=0]{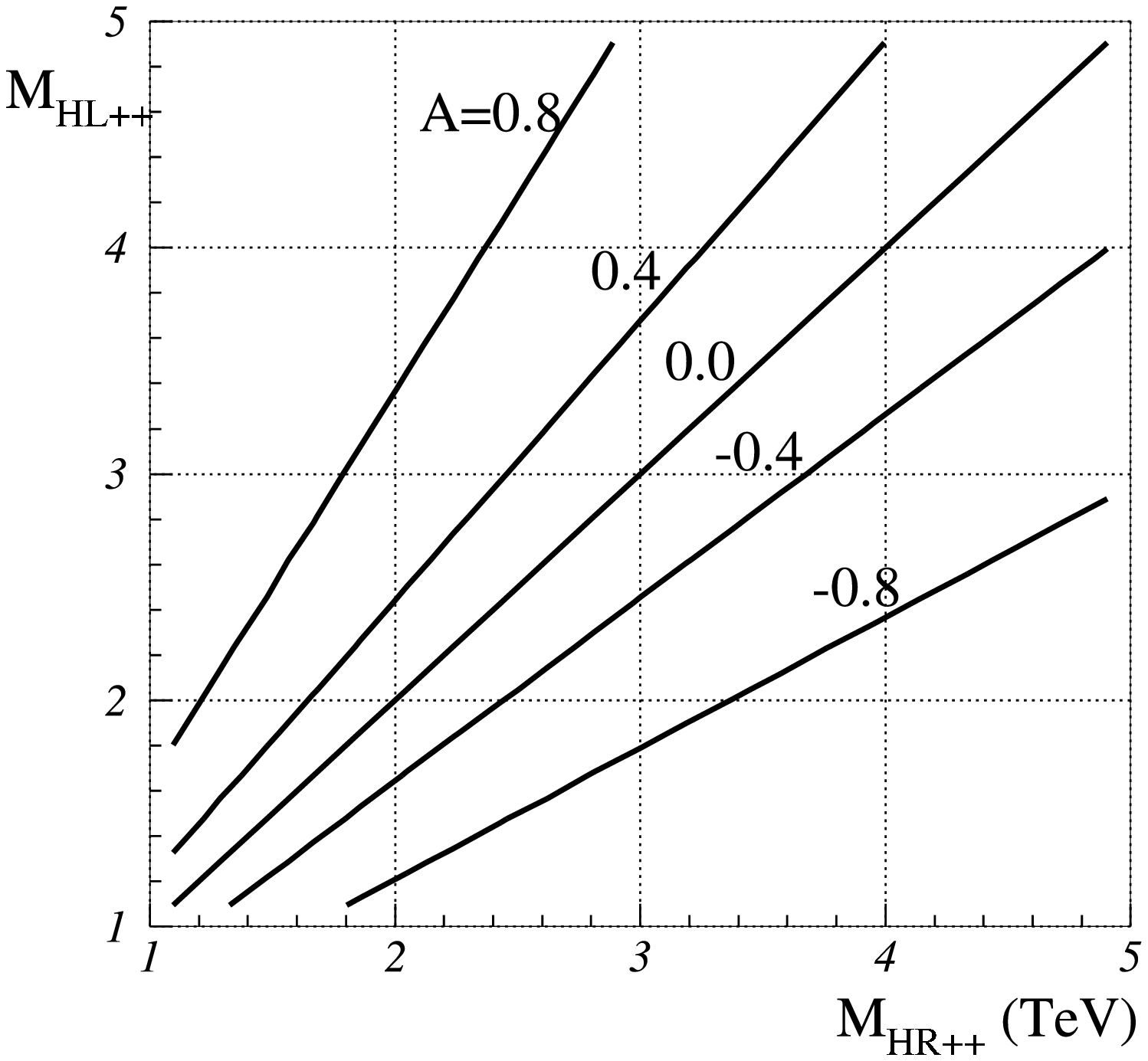} 
\includegraphics[width=6.5cm,angle=0]{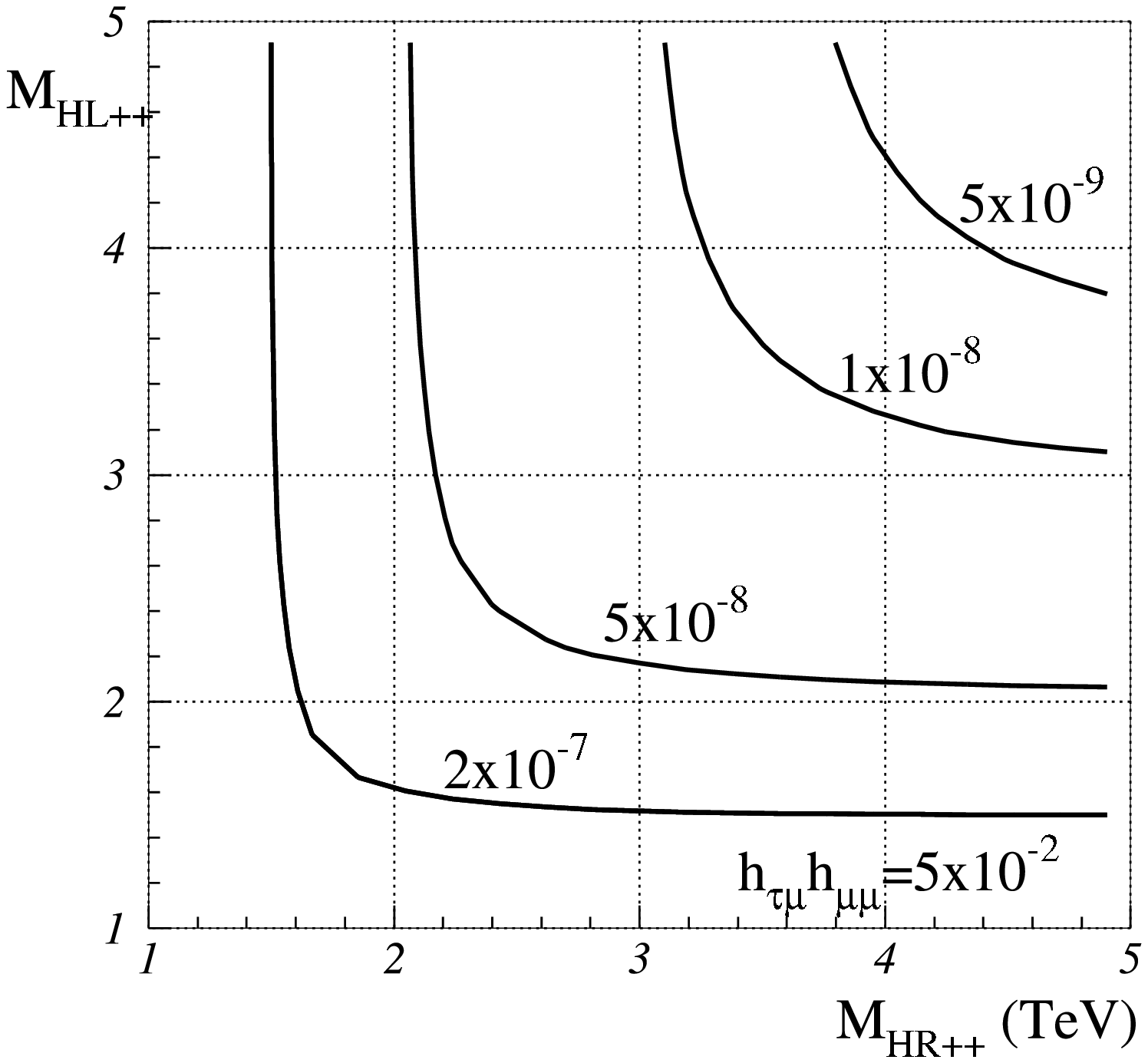}
\end{center}
\caption{%
(a)  ${\cal A}(\tau\to lll)$ and (b)  BR($\tau\to \mu\mu\mu$) and in the plane
 ($M_{H^{\pm\pm}_L}$, $M_{H^{\pm\pm}_R}$)
}
\label{asym_br}
\end{figure}
The expressions for $g_3$ and $g_4$ in the three models under
consideration are given in Table~2.
In the manifest LR symmetric model
$h_{ij}$ cancels out in the expression for ${\cal A}(\tau\to lll)$ 
leaving a simple dependence on $M_{H^{\pm\pm}_L}$ and $M_{H^{\pm\pm}_R}$ 
which applies
to all six $\tau\to lll$ decays:
\begin{equation}
{\cal A}(\tau\to lll)=\frac{1/M_{H^{\pm\pm}_R}^4-1/M_{H^{\pm\pm}_L}^4}{1/M_{H^{\pm\pm}_R}^4+1/M_{H^{\pm\pm}_L}^4} \ .
\end{equation}
In Fig.\ref{asym_br} (a) ${\cal A}(\tau\to lll)$ in plotted in the plane
($M_{H^{\pm\pm}_L}, M_{H^{\pm\pm}_R}$).
Clearly the case of degeneracy ($M_{H^{\pm\pm}_L}=M_{H^{\pm\pm}_R}$)
gives ${\cal A}(\tau\to lll)=0$, while $M_{H^{\pm\pm}_L} > M_{H^{\pm\pm}_R}$
($M_{H^{\pm\pm}_L} < M_{H^{\pm\pm}_R}$)
results in positive (negative) ${\cal A}(\tau\to lll)$.
For the HTM and ZBM the asymmetry is maximal, 
being $-1$ and $+1$ respectively. Consequently,
${\cal A}(\tau\to lll)$ in the
LR symmetric model may differ significantly from the
corresponding value in
models with only a $H^{\pm\pm}_L$ (HTM) or $H^{\pm\pm}_R$ (ZBM). 
Thus ${\cal A}(\tau\to lll)$
has the potential to discriminate between the various
models with a $H^{\pm\pm}$.
In addition, in the context of the
LR model a measurement of ${\cal A}(\tau\to lll)$ provides important 
information on the ratio $M_{H^{\pm\pm}_L}/M_{H^{\pm\pm}_R}$, which could
assist the direct searches for $H^{\pm\pm}_L$ and $H^{\pm\pm}_R$ 
at high energy colliders. For comparison we show in 
Fig.\ref{asym_br} (b) the dependence of BR$(\tau\to \mu\mu\mu)$
on $M_{H^{\pm\pm}_L}$ and $M_{H^{\pm\pm}_R}$ for 
$|h_{\mu\mu}h_{\tau\mu}^\ast |=0.05$.

\begin{figure}[ht]
\centering
\begin{picture}(200,200)  
\put(45,75){\makebox(100,20){\epsfig{figure=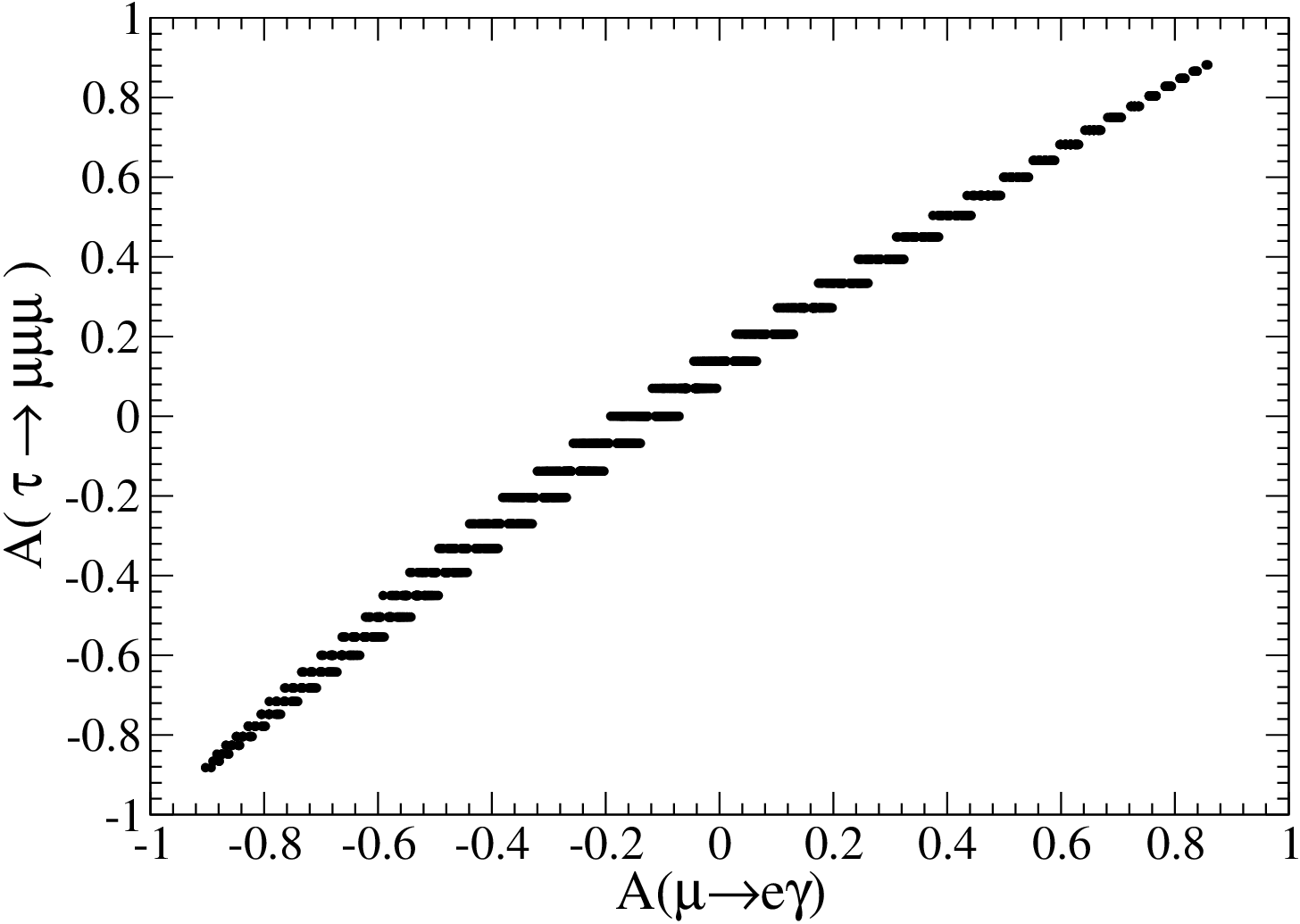,width=8cm}}}
\end{picture}~~
\caption{Correlation between ${\cal A}(\mu\to e \gamma)$ and ${\cal A}(\tau \to \mu\mu\mu)$
in the LR model.}
\label{LRmodel_asym}
\end{figure}

Of the order of 50 $\tau\to lll$ events would
be needed to distinguish ${\cal A}(\tau\to lll)=+1$ from $-1$.  
A high luminosity upgrade of the existing B factories
anticipates up $10^{10}$ $\tau^+\tau^-$ pairs and thus 
BR$(\tau\to lll)>10^{-8}$ would allow measurements of
${\cal A}(\tau\to lll)$.
From Fig.\ref{asym_br} (a) it is clear that a LR model 
with $M_{H^{\pm\pm}_L}\ll M_{H^{\pm\pm}_R}$
($M_{H^{\pm\pm}_R}\ll M_{H^{\pm\pm}_L}$)
will give an almost maximal ${\cal A}(\tau\to lll)$ and
consequently would be difficult to distinguish from the HTM (ZBM). 
However, if a signal were also observed for 
$\mu^+\to e^+\gamma$, the analogous $P$ odd asymmetry, 
${\cal A}(\mu\to e\gamma$),
can serve as an additional discriminator:
\begin{equation}
{\cal A}(\mu\to e\gamma)=\frac{|A_L|^2-|A_R|^2}{|A_L|^2+|A_R|^2} \ .
\end{equation}
In CASE I, Fig.\ref{fig:3} there is a parameter space 
for BR($\tau\to \mu\mu\mu)\approx 10^{-8}$ and BR($\mu\to e\gamma)\approx
10^{-12}$), which might provide sufficient events for both
asymmetries to be measured.
In contrast to $\tau\to \mu\mu\mu$, the loop induced
decay $\mu\to e\gamma$ can be mediated by
$H^{\pm\pm}_{L,R}$, $H^\pm_1$ and $W^\pm_R$ in the LR symmetric model.
One may write a simplified formula for $A_L$ and $A_R$
(written explicitly in Eqs.~(\ref{mu_ey_AL}) and (\ref{mu_ey_AR})), where
$a,b,c,d$ are functions of masses and the heavy 
neutrino mixing matrix $K_R$: 
\begin{eqnarray}
A_L&=&a~(M_{H^{\pm\pm}_R},K_R)+ b~(M_{W^\pm_R},K_R,M_i) \ , \nonumber \\
A_R&=& c~(M_{H^{\pm\pm}_L},K_R)+d~(M_{H^{\pm}_1},K_R) \ .
\end{eqnarray}
In the LR model usually $a,c \gg  b,d$ and so the dominant contribution
to ${\cal A}(\mu\to e\gamma)$ arises from $H^{\pm\pm}_L$ and 
$H^{\pm\pm}_R$.
%$H^\pm_L$ contribution is suppressed relative to
%the $H^{\pm\pm}_L$ by a factor ($1/8$). Although this could
%be compensated by $M_{H^\pm}<M_{H^\pm H^\pm}$, in the LR
%the condition $M_{H^\pm_L}\approx M_{H^\pm_L H^\pm_L}$ ensures that
%the contribution from 
%$H^\pm_L$ is always subdominant. The $W_R$ contribution is 
%also suppressed relative to $H^{\pm\pm}_R$. 
Hence in LR models one expects ${\cal A}(\mu\to e\gamma)\sim %\approx 
{\cal A}(\tau\to lll$). 
%to within $5\to 10\%$. 
This is shown in Fig.\ref{LRmodel_asym} where
${\cal A}(\tau\to \mu\mu\mu$) is plotted against ${\cal A}(\mu\to e\gamma)$ 
for CASE I with $\sin \theta_{13}=0.2$ and $\delta=\pi$. 
The plotted points correspond to observable rates for both
LFV decays (taken to be $10^{-9} \le$ BR($\tau \to \mu\mu\mu) \le 10^{-7}$ and 
$10^{-14} \le$ BR($\mu \to e\gamma) \le 10^{-11}$),
within the following parameter region:
$M_{W_2}=3$ TeV, 1 TeV $\leq M_i \leq 5$ TeV, 2 TeV $\leq 
M_{H^{\pm\pm}_L}=M_{H^\pm_1}\ne M_{H^{\pm\pm}_R} \leq$ 4 TeV.
Each point also satisfies the constraint BR$(\mu\to eee) < 10^{-12}$.
Clearly the vast majority of the points are close to the
line ${\cal A}(\mu\to e\gamma)={\cal A}(\tau\to lll)
$, showing that
the diagrams involving $H^{\pm\pm}_L$ and $H^{\pm\pm}_R$
give the dominant contribution over most of the
parameter space. The asymmetries differ sizeably only when
$M_i$ and $M_{W_R}$ are considerably smaller than
$M_{H^{\pm\pm}_L}$ and $M_{H^{\pm\pm}_R}$.
In the HTM one has:
\begin{eqnarray}
A_L&=&0 \ ,\nonumber \\
A_R&=&c~(M_{H^{\pm\pm}_L},h_{ij})+d~(M_{H^{\pm}_L},h_{ij}) \ .
\end{eqnarray}
and thus ${\cal A}(\mu\to e\gamma$) is maximal.
In the ZBM:
\begin{eqnarray}
A_L&=&a~(M_{H^{\pm\pm}_R},h_{ij}) \ , \nonumber \\
A_R&=&d~(M_{H^{\pm}_L},f_{ij}) \ .
\end{eqnarray}
${\cal A}(\mu\to e\gamma$) may take any value 
since the masses of $H^\pm_L$ and $H^{\pm\pm}_R$
are unrelated and $h_{ij}\ne f_{ij}$ in general.
The allowed ranges of ${\cal A}(\tau\to lll)$ and 
${\cal A}(\mu\to e\gamma)$
in the three models under consideration are summarized in Table~3.
It is clear that if signals for both 
$\tau\to lll$ and $\mu\to e\gamma$ are observed,
the corresponding asymmetries may act as a powerful 
discriminator of the models.

\begin{table}
\begin{center}
\caption{$P$ odd asymmetries ${\cal A}(\tau\to lll)$
and ${\cal A}(\mu\to e\gamma)$ in the three models}
%\begin{minipage}[t]{0.8\textwidth}
\begin{tabular}{cccc}
\hline
 & HTM ($H^{\pm\pm}_L$) & ZBM ($H^{\pm\pm}_R$) &  LR ($H^{\pm\pm}_{L,R}$) \\
\hline
${\cal A}(\mu \to e \gamma$) & $-1$ & $- 1< {\cal A} < +1$ &   $ -1< {\cal A} < +1$    \\
${\cal A}(\tau\to lll$)  & $-1$  & +1 &  $ -1< {\cal A} < +1$  \\
\hline
\end{tabular}
%\end{minipage}
\end{center}
\end{table}

%\begin{figure}[h]
%\begin{center}
%\includegraphics[width=8.5cm,angle=0]{muey.ps} 
%\end{center}
%\caption{%
%${\cal A}(\mu\to e\gamma$) in the plane
%($M_{W_R}$, $M_{N_1}$), taking 
%$M_{H^{\pm\pm}_L}=M_{H^{\pm\pm}_R}=2000$ GeV.
%}
%\label{fig1}
%\end{figure}

\section{Conclusions}
The Left-Right symmetric extension of the
Standard Model with TeV scale breaking
of $SU(2)_R$ via a right handed Higgs isospin triplet
vacuum expectation value provides an attractive explanation for  
neutrino masses via the seesaw mechanism.
The doubly charged scalars $H^{\pm\pm}_L$ and $H^{\pm\pm}_R$ 
with mass of order TeV mediate the LFV decays $\tau \to lll$ 
at tree-level via a Yukawa coupling $h_{ij}$
which is related to the  Maki-Nakagawa-Sakata matrix $(V_{\rm MNS})$. 
We introduced four ansatz for the origin of the
bi-large mixing in $V_{\rm MNS}$ which
satisfy the stringent bound BR($\mu\to eee)<10^{-12}$ in distinct ways.
A numerical study of the magnitude 
and correlation of BR($\tau^\pm \to lll$) and BR($\mu\to e\gamma$)
was performed. It was shown that the number of observable rates 
for such LFV decays depends sensitively on the origin of 
the bi-large mixing in $V_{\rm MNS}$, with multiple LFV signals being
possible in specific cases.

If a signal for $\tau \to lll$ were observed we showed how 
the definition of an angular asymmetry provides
information on the relative strength 
of the contributions from $H^{\pm\pm}_L$ and $H^{\pm\pm}_R$.
Such an asymmetry may also be used to distinguish
the LR symmetric model from other models 
which contain either $H^{\pm\pm}_L$ or $H^{\pm\pm}_R$
and thus predict maximal asymmetries.

\section*{Acknowledgements}
Y.O was supported in part by the Grant-in-Aid for Science
Research, Ministry of Education, Science and Culture, Nos.
16081211, 13135225 and 17540286.
A.G.A was supported by National Cheng Kung University grant
OUA 95-3-2-057.

\end{document}